\documentclass[11pt,english]{article}
\usepackage{lmodern}

\usepackage[T1]{fontenc}
\usepackage[latin9]{inputenc}
\usepackage[letterpaper]{geometry}
\usepackage{color}
\usepackage{amsmath}
\usepackage{amssymb}
\usepackage[authoryear]{natbib}
\usepackage{subfigure}
\usepackage{graphicx}

\usepackage{lmodern}
\usepackage[T1]{fontenc}
\usepackage[latin9]{inputenc}
\usepackage[letterpaper]{geometry}
\usepackage{color}

 \usepackage{float}

\usepackage{textcomp}

\usepackage[authoryear]{natbib}

\usepackage{amsmath}
\usepackage{amssymb}
\usepackage{amsfonts}
\usepackage{multirow}
\usepackage{array,epsfig,rotating}
\usepackage[]{hyperref}  

\usepackage{graphicx}
\usepackage{amsthm}
\usepackage{mathtools}
\usepackage{lscape}
\usepackage{xcolor}
\usepackage{tabularx}
\usepackage{url}
\usepackage{pdflscape}
\usepackage{bbm}
\usepackage{lscape}
\usepackage{bbm}
\usepackage{latexsym}
\usepackage{mathrsfs}
\usepackage{setspace}

\makeatletter

\@ifundefined{definecolor}
 {\usepackage{color}}{}
\@ifundefined{definecolor}{\@ifundefined{definecolor}
 {\usepackage{color}}{}
}{}

\newtheorem{theorem}{Theorem}\newtheorem{lemma}{Lemma}\newtheorem{corollary}{Corollary}
\newtheorem{remark}{Remark}\newtheorem{assum}{Assumption}

\newcommand{\lyxmathsym}[1]{\ifmmode\begingroup\def\b@ld{bold}
  \text{\ifx\math@version\b@ld\bfseries\fi#1}\endgroup\else#1\fi}\makeatother

\usepackage{babel}

\theoremstyle{definition}

\theoremstyle{remark}

\numberwithin{equation}{section}

\renewcommand{\vec}{\mathrm{vec}}
\newcommand{\vech}{\mathrm{vech}}
\newcommand{\var}{\mathrm{var}}

\newcommand{\rank}{\mathrm{rank}}
\newcommand{\tr}{\mathrm{tr}}
\makeatother

\usepackage{babel}

\begin{document}
\global\long\def\baselinestretch{1.5}
 \fontsize{12}{14pt plus.8pt minus .6pt}\selectfont \vspace{0.8pc}

\setcounter{footnote}{1}
\title{The likelihood ratio test for structural changes in factor
models\footnote{
We would like to thank editors Elie Tamer and Serena Ng, and two anonymous referees for their insightful suggestions. We extend our thanks to the participants of the 2022 NBER/NSF Time Series Conference at Boston University.
Jiangtao Duan's research is partially supported by National Nature Science Foundation of China (No.12101146, 12371263).}}
\date{\vspace{-15ex}}
\maketitle

\vspace{.4cm} \centerline{ Jushan Bai\textsuperscript{1}, Jiangtao Duan\textsuperscript{2}, Xu Han\textsuperscript{3}  } \vspace{.4cm} \centerline{\it
\textsuperscript{1}Columbia University, \textsuperscript{2}Xidian University and \textsuperscript{3}City University of Hong Kong}

\renewcommand{\thefootnote}{\arabic{footnote}}
\setcounter{footnote}{0}

\footnotetext[1]{Department of Economics, Columbia University. E-mail: jb3064@columbia.edu}

\footnotetext[2]{School of Mathematics and Statistics, Xidian University. E-mail: duanjiangtao@xidian.edu.cn}

\footnotetext[3]{Department of Economics and Finance, City University of Hong Kong. E-mail: xuhan25@cityu.edu.hk}

\begin{quotation}
\noindent \textit{\large Abstract:}{\large \par}

A factor model with a break in its factor loadings is observationally equivalent to a model without changes in the loadings but with a change in the variance of its factors. This approach effectively transforms a high-dimensional structural change problem into a low-dimensional problem. This paper considers the likelihood ratio (LR) test for a variance change in the estimated factors. The LR test implicitly explores a special feature of the estimated factors: the pre-break and post-break variances can be a singular matrix under the alternative hypothesis, making the LR test diverging faster and thus more powerful than Wald-type tests. The better power property of the LR test is also confirmed by simulations. We also consider mean changes and multiple breaks. We apply this procedure to the factor modeling of the US employment and study the structural change problem
using monthly industry-level data.

\textit{\large \vspace{9pt}
 }{\large \par}
 \vspace{9pt}
\par
\end{quotation}
\textit{\large \vspace{-1em}
 }{\large \par}

\newpage

\section{Introduction}

Factor models are effective tools for summarizing information in large datasets and are widely used
in economics and finance. Examples include diffusion index forecasting (Stock and Watson, 2002, 2009), asset pricing (Ross, 1976; Fama and French, 1992; Feng et al., 2020), and macroeconomic policy evaluation (e.g., Bernanke et al., 2005; Han, 2018).
Structural change is a common
phenomenon in economic variables and is more likely to occur in
high-dimensional data. To better understand the
data structure and ensure the validity of subsequent analyses,
it is useful to check the structural stability of factor models.

This paper focuses on testing structural breaks in the factor
loading matrix. It is known that a factor model with structural
breaks in its factor loading matrix is observationally equivalent
to a model with time-invariant loadings but a potentially larger number of factors (hereinafter referred to as ``pseduo-factors'') than the original model. This approach effectively translates
the original and challenging high-dimensional testing
problem into a low-dimensional problem. Based on this
fact, the literature proposes numerous procedures to examine the moments
of the pseudo-factors. For example, Chen, Dolado, and Gonzalo (2014) (CDG hereafter)
propose Wald and Lagrange multiplier (LM) tests to examine the coefficients in the regression
of the first estimated factor on the remaining factors. \citet{Han2015}
(HI hereafter) develop Wald and LM tests that compare the pre- and
post-break second moments of the estimated factors. Baltagi et al. (2021)
generalize HI's results and propose tests for multiple breaks in
the factor loading matrix. Although these tests are consistent under
certain alternative hypotheses, simulation evidence shows that they
may not be powerful for moderate breaks in finite samples.

We consider a (quasi-)LR test.
Our research is motivated by
Duan, Bai, and Han (2022, DBH hereafter), who show that the quasi-maximum likelihood
(QML) estimator of the break point is consistent when the subsample
covariances of the pseudo-factors are singular. By consistency, we mean that the probability that the estimated break date is exactly equal to the true break date approaches one as the sample size grows. This implies a faster rate than the usual $T$-consistency in terms of break fractions.

By construction, the LR test
statistic is equal to the likelihood function evaluated at the estimated break point estimator. Thus, the faster convergence of the QML break estimator (under the alternative) is expected to correspond to a more powerful test
of the LR statistic. In contrast,
the sup-Wald and sup-LM statistics of CDG (2014) and HI do not generate consistent estimators for the break point
under the alternative. Therefore,
these tests are less powerful than the LR statistic.

LR tests are studied by Qu and Perron (2007) and Perron, Yamamoto, and Zhou (2020) for observable variables. In factor models, because the factor process is unobservable and is estimated subject to normalization restrictions, the limiting distributions of our LR tests are different from those in Qu and Perron (2007), except for the case of testing a single change. The limiting distribution depends on the increments of the Brownian bridge instead of Brownian motion. Interestingly, the normalization restrictions make the derivation of the limiting distribution even easier. The LR test is asymptotically equivalent to HI's sup-Wald test under the null hypothesis.
But its behavior under the alternative is more difficult to analyze for factor models. Inspired by the results in DBH (2022), we show that the LR test is diverging faster than the Wald test under the alternative.
The higher power of our LR test is related to the insight that using multiple time series helps
identify a break point (Bai et al., 1998). This insight culminates in the finding in \cite{Bai2010} that it is possible to precisely identify the break point for common mean and variance breaks in large panels. Under the latent factor
setup, the large cross-section still plays an important role, as it helps enforce the singularity for
subsample pseudo-factor variance as $N$ grows, which is responsible for the superior power and accurate
estimation of the break point under the alternative. \cite{Bai2010} further show that even in the absence of a
change in variance, QML allowing a variance change provides a more efficient estimator of the break point
(in mean) than imposing the correction assumption of no change in variance. Taken together, these
results support our use of the likelihood approach for testing structural changes.

Our paper is also
related to, but substantially different from, other studies examining
structural changes in factor models. The earlier literature focuses
on the low-dimensional testing problem of whether the factor loadings of
an individual variable have structural changes (e.g., Stock and Watson, 2009;
Breitung and Eickmeier, 2011; and Yamamoto and Tanaka, 2015). In comparison, we are
interested in testing changes in the entire loading matrix. Cheng et al. (2016) develop
a Lasso estimator to determine whether there is a change in the factor
loading matrix or the number of factors. While they concentrate on consistent model selection, our emphasis is on hypothesis testing. Su and Wang (2020) propose a test for identifying smooth changes in factor loadings using local principal components, while our test employs a quasi-maximum likelihood approach. Another related area of research focuses on the consistent estimation of factors under time-varying factor loadings. For example,
Bates et al. (2013) established conditions under which the time variation of factor loadings can be ignored. However, our
framework violates Bates et al.'s conditions because a significant proportion of factor loadings undergo a break at a common time
under the alternative hypothesis. Mikkelsen et al. (2019) assumed factor loadings following stationary VAR processes, but our
framework's factor loadings experience a one-time shift, which means they are not stationary over time. Therefore, their method
cannot be applied to our setting, unlike their setting where the time variation in factor loadings can be absorbed as idiosyncratic
components.

Our work is further related to the literature on break point estimation for large factor models (e.g.,
Chen, 2015; Cheng et al., 2016; Baltagi et al., 2017; Barigozzi et al., 2018; Ma and Su, 2018; Bai
et al., 2020). Most of the research is not likelihood-based, with the exception of DBH (2022), as explained above.
Bai (2000) study the QML estimation of mean and variance breaks, but in a low-dimensional
vector autoregressive setting. In addition, there is a large literature that tests breaks in a traditional
time series setting, such as Andrews (1993), Bai and Perron (1998) and Bergamelli et al.
(2019), among others.

The rest of this paper is organized as follows. Section 2 introduces
the representation of the factor model with a single break in the
factor loading matrix. Section 3 describes the LR test, establishing
its asymptotic distribution under the null hypothesis and furthering deriving the
divergence rate of the statistic under the alternative hypothesis. Section
4 generalizes the LR test to allow for factor mean changes and multiple
breaks in the loading matrix. Section 5 investigates the finite-sample
properties of the LR procedure via simulations. Section 6 implements the LR test for monthly US industry-level employment data.

The following notations are used. For an
$m\times n$ real matrix $\mathbb{A}$, we denote its Frobenius norm
as $\|\mathbb{A}\|=[tr(\mathbb{A}\mathbb{A}^{\prime})]^{1/2}$. For
a real number $x$, $[x]$ represents the integer part of $x$. $|\mathcal{A}|$
denotes the cardinality of a set $\mathcal{A}$. vech$(\cdot)$ is equal to the column-wise vectorization of a square matrix with the upper triangular excluded.

\section{The factor model and hypotheses}

Consider the following factor model that allows for a structural break
in the factor loading matrix: \begin{eqnarray}
x_{t}=\begin{cases}
\Lambda_{1}f_{t}+e_{t}\  & \mathrm{for}\ t=1,2,\cdots,T_{1}\\
\Lambda_{2}f_{t}+e_{t}\  & \mathrm{for}\ t=T_{1}+1,\cdots,T,\end{cases}\label{model_vector0}\end{eqnarray}
where $x_{t}=[x_{1t},\cdots,x_{Nt}]^{'}$ is the $N$-dimensional demeaned
observation at time $t$; $f_{t}$ is an $r-$dimensional vector of
unobserved common factors and $E(f_{t}f_{t}^{\prime})=\Sigma_{F}$;
$T_{1}$ is an unknown break date; $\Lambda_{1}=[\lambda_{11},\cdots,\lambda_{N1}]^{'}$
and $\Lambda_{2}=[\lambda_{12},\cdots,\lambda_{N2}]^{'}$ are the
$N\times r$ pre- and post-break factor loadings, respectively; and
$e_{t}=[e_{1t},\cdots,e_{Nt}]^{'}$ is the $N\times1$ idiosyncratic
error that may have serial and cross-sectional dependence along with heteroskedasticity. We define $\pi_{1}=T_{1}/T\in(0,1)$
as the break fraction, which is assumed to be a fixed constant. This
implies that $T_{1}$ is a sequence that depends on $T$.  For notational simplicity, we suppress
the dependence of $T_{1}$ on $T$.

We are interested in testing the null hypothesis of no structural
break in the factor loadings, i.e., \begin{eqnarray}
 &  & \mathbb{H}_{0}:\,\lambda_{i1}=\lambda_{i2}\ \forall i.\label{null_test}\end{eqnarray}
against the alternative hypothesis that a non-negligible portion of
the cross sections have a break in their loadings at a common time,
i.e., \begin{eqnarray}
 &  & \mathbb{H}_{1}:\,\lambda_{i1}\neq\lambda_{i2}\ \mathrm{for}\ i\in\mathcal{J}\label{alternative_test}\end{eqnarray}
where $|\mathcal{J}|/N\to b_{0}\in(0,1]$ as $N\to\infty$.

Under $\mathbb{H}_{0}$, (\ref{model_vector0}) is a standard factor
model with time-invariant factor loadings and $r$ denotes the number
of original factors. Under $\mathbb{H}_{1}$, it is well known that
the factor model is observationally equivalent to a model with time-invariant
loadings and potentially more pseudo-factors (e.g., HI, 2015;
Baltagi et al., 2017). To capture the factor dimension augmentation
caused by the break, we follow the framework of DBH (2022)
and set $r$ as the number of pseudo-factors in (\ref{model_vector0}).
We set \[
\Lambda_{1}=\Lambda B,\ \ \Lambda_{2}=\Lambda C,\]
where $\Lambda$ is an $N\times r$ matrix with full column rank $r$,
$B$ and $C$ are $r\times r$ matrices, $\rank(B)=r_{1}\le r$,
and $\rank(C)=r_{2}\le r$.

For a given split point $k$, define \begin{align*}
X_{k}^{(1)} & =[x_{1},...,x_{k}]^{\prime},\ X_{k}^{(2)}=[x_{k+1},...,x_{T}]^{\prime},\ F_{k}^{(1)}=[f_{1},...,f_{k}]^{\prime},\ F_{k}^{(2)}=[f_{k+1},...,f_{T}]^{\prime},\\
e_{k}^{(1)} & =[e_{1},...,e_{k}]^{\prime},\ e_{k}^{(2)}=[e_{k+1},...,e_{T}]^{\prime}.\end{align*}
Thus, (\ref{model_vector0}) can be rewritten in the following matrix
format:\begin{eqnarray}
\left[\begin{array}{c}
X_{T_{1}}^{(1)}\\
X_{T_{1}}^{(2)}\end{array}\right] & = & \left[\begin{matrix}F_{T_{1}}^{(1)}\Lambda_{1}^{\prime}\\
F_{T_{1}}^{(2)}\Lambda_{2}^{\prime}\end{matrix}\right]+\left[\begin{array}{c}
e_{T_{1}}^{(1)}\\
e_{T_{1}}^{(2)}\end{array}\right]=\left[\begin{matrix}F_{T_{1}}^{(1)}(\Lambda B)^{\prime}\\
F_{T_{1}}^{(2)}(\Lambda C)^{\prime}\end{matrix}\right]+\left[\begin{array}{c}
e_{T_{1}}^{(1)}\\
e_{T_{1}}^{(2)}\end{array}\right],\nonumber \\
 & = & \left[\begin{matrix}F_{T_{1}}^{(1)}B^{\prime}\\
F_{T_{1}}^{(2)}C^{\prime}\end{matrix}\right]\Lambda^{\prime}+\left[\begin{array}{c}
e_{T_{1}}^{(1)}\\
e_{T_{1}}^{(2)}\end{array}\right],\nonumber \\
 & = & G\Lambda^{\prime}+e,\label{DBH}\end{eqnarray}
which is an observationally equivalent representation with a time-invariant
loading matrix $\Lambda$ and pseudo-factors $G$ with $\rank(G)=r$.

The representation in (\ref{DBH}) is flexible for both
the null and alternative hypotheses. $r_{1}=\rank(B)$ and
$r_{2}=\rank(C)$ are the numbers of factors before and after
the break. Accordingly, under $\mathbb{H}_{0}$ we can set $B=C=I_{r}$,
and pseudo-factors $G$ coincide with the original factors. Under
$\mathbb{H}_{1}$, we can incorporate different types of changes by
controlling the ranks of $B$ and $C$. Following DBH (2022), loading
changes can be divided into three types: Type (1), in which both $B$ and $C$
are singular (i.e., both $r_{1}$ and $r_{2}$ are less than $r$,
so (\ref{DBH}) can capture the factor dimension enlargement
caused by the break); Type (2), in which only $B$ or $C$ is singular (emerging
or disappearing factors); and Type (3), in which both $B$ and $C$ are nonsingular
(rotational change in loadings). In practice, Types (1) and (2) are more common than Type (3).

When $H_{1}$ is true, the pre- and post-break second moments of the pseudo-factors
are $B\Sigma_{F}B^{\prime}$ and $C\Sigma_{F}C^{\prime}$, respectively.
Thus, various tests (e.g., the sup-Wald and sup-LM type statistics developed
by \citet{chen2014} and \citet{Han2015}) propose to compare the
subsample second moments of the factors under the assumption that
$\Sigma_{F}$ is constant over time. Although these tests are consistent
under the alternative hypothesis (e.g., Han and Inoue, 2015), simulation
evidence shows that they have limited power in small samples when
the break size is moderate.

To develop a more powerful test, we take a different
route. Instead of comparing the variance of the factors before and
after the break, we construct the maximum of LRs (sup-LR)
 based on the quasi-likelihood functions of the factors evaluated at different potential break points. The proposed
LR test is expected to be more powerful than the sup-Wald type test because the QML estimator for the break point under the alternative is more precise than the estimators based on the least squares estimators (the maximum of Wald-type estimators).
DBH (2022) show that the
break point estimated by QML is equal to
$T_{1}$, with a probability approaching one as $N,T\to\infty$ when
$B$ or $C$ or both are singular. The Wald-type estimators do not have this property and are only
stochastically bounded around $T_1$.

\section{Likelihood ratio test}

Because $G$ is unobserved, we need an estimator of $G$ to construct
our test statistic. Let $\hat{G}=[\hat{g}_{1},\cdots,\hat{g}_{T}]^{\prime}$
denote the principal components (PC) estimator of $G$ under the usual
identification condition:\begin{equation}
T^{-1}\hat{G}^{\prime}\hat{G}=I_{r}\;\text{and}\;\hat{\Lambda}^{\prime}\hat{\Lambda}\text{ is a diagonal matrix.}\label{restrictions}\end{equation}
Thus, $\hat{G}$ is $\sqrt{T}$ times the eigenvectors corresponding to the $r$ largest eigenvalues of the
$T\times T$ matrix  $X'X$. Because the data are demeaned,
the PC estimator satisfies $T^{-1}\sum_{t=1}^{T}\hat{g}_{t}=0$.
For a given split point $k$, we estimate the pre-$k$ and post-$k$
factor variances by \[
\hat{\Sigma}_{1}(k)=\frac{1}{k}\sum\limits _{t=1}^{k}\hat{g}_{t}\hat{g}_{t}^{\prime},\;\hat{\Sigma}_{2}(k)=\frac{1}{T-k}\sum\limits _{t=k+1}^{T}\hat{g}_{t}\hat{g}_{t}^{\prime}.\]
Here, we do not subtract the subsample mean, which is considered later.
The quasi-Gaussian likelihood for a break point at $k$ is given by \begin{equation}
\mathcal{L}(k)=k\log(|\hat{\Sigma}_{1}(k)|)+(T-k)\log(|\hat{\Sigma}_{2}(k)|).\label{eq:Likelihood}\end{equation}
Because the sample variance of $\hat{g}_{t}$ is always equal to an
identity by (\ref{restrictions}), the log-likelihood of no change for the entire sample
is \[
\mathcal{L}_{0}=T \log (|\frac 1 T \sum_{t=1}^T \hat g_t \hat g_t'|) = T\log(|I_{r}|)\equiv 0.\]
Thus, the LR for testing the null hypothesis of no change against the alternative of a change occurring at
$k$ is $LR(k):=\mathcal{L}_0-\mathcal{L}(k)$, which can be written as

\[
LR(k)\equiv-k\log(|\hat{\Sigma}_{1}(k)|)-(T-k)\log(|\hat{\Sigma}_{2}(k)|).\]
To test against (\ref{alternative_test}) for an unknown
$T_{1}$, we employ the supreme LR statistic\begin{equation}
\sup-LR\equiv\sup\limits _{[\epsilon T]\le k\le[(1-\epsilon)T]}LR(k),\label{LR_test_hatG}\end{equation}
where $\epsilon\in(0,0.5)$.

The identification condition (3.1) is chosen for computational convenience, but the LR
test can be employed under any valid identification condition. The explanation is provided in the Appendix.

\subsection{Assumptions}

To study the limiting null distribution of the LR test statistics,
we make the following standard assumptions for the approximate factor
model that allow for the functional central limit theorem.
\begin{assum}\label{factors} $E\|f_{t}\|^{4}<M<\infty$,
$E(f_{t}f_{t}^{\prime})=\Sigma_{F}$, where $\Sigma_{F}$ is positive
definite, and $T^{-1}\sum_{t=1}^{[\pi T]}f_{t}f_{t}^{\prime}\to_{p}\pi\Sigma_{F}$
uniformly for $\pi\in[0,1]$.
\end{assum}
\begin{assum}\label{Factor_Loadings} Let $\lambda_{i}^{\prime}$
be the $i$-th row of $\Lambda$. $\|\lambda_{i}\|\leq\bar{\lambda}<\infty$
for $i=1,\cdots,N$, $\|N^{-1}\Lambda^{\prime}\Lambda-\Sigma_{\Lambda}\|\rightarrow0$
for some $r\times r$ positive definite matrix $\Sigma_{\Lambda}$.
\end{assum}
\begin{assum}\label{Depen_and_Hetero} There exists a positive
constant $M<\infty$ such that\par \begin{itemize}

\item[(i)] $E(e_{it})=0$ and $E|e_{it}|^{8}\leq M$ for all $i=1,\cdots,N$
and $t=1,\cdots,T$;

\item[(ii)] $E(e_{s}^{\prime}e_{t}/N)=E(N^{-1}\sum_{i=1}^{N}e_{is}e_{it})=\gamma_{N}(s,t)$
and $\sum_{s=1}^{T}|\gamma_{N}(s,t)|\leq M$ for every $t\leq T$;

\item[(iii)] $E(e_{it}e_{jt})=\tau_{ij,t}$ with $|\tau_{ij,t}|<\tau_{ij}$
for some $\tau_{ij}$ and for all $t=1,\cdots,T$ and $\sum_{j=1}^{N}|\tau_{ij}|\leq M$
for every $i\leq N$;

\item[(iv)] $E(e_{it}e_{js})=\tau_{ij,ts}$, $
\frac{1}{NT}\sum\limits _{i,j,t,s=1}|\tau_{ij,ts}|\leq M;$ and

\item[(v)] For every $(s,t)$, $E\left|N^{-1/2}\sum_{i=1}^{N}(e_{is}e_{it}-E[e_{is}e_{it}])\right|^{4}\leq M$.
\end{itemize} \end{assum}
\begin{assum}\label{Weak_Dependence}\par There exists
a positive constant $M<\infty$ such that
$E\left(\frac{1}{N}\sum\limits _{i=1}^{N}\left\Vert \frac{1}{\sqrt{T}}\sum\limits _{t=1}^{T}f_{t}e_{it}\right\Vert ^{2}\right)\leq M.$
 \par \end{assum}
\begin{assum}\label{eigenvalues} The eigenvalues of $\Sigma_{G}\Sigma_{\Lambda}$
are distinct where $\Sigma_{G}=\pi_1 B\Sigma_{F}B^{\prime}+(1-\pi_1)C\Sigma_{F}C^{\prime}$.\end{assum}
\noindent \begin{assum}\label{factor_error} There exists an $M<\infty$
such that for all $N$ and $T$ and for each $t$,\\
(i)\;$E\left(  \left\Vert \frac{1}{\sqrt{NT}}\sum_{s=1}^{T}\sum_{k=1}^{N}f_{s}[e_{ks}e_{kt}-E(e_{ks}e_{kt})]\right\Vert ^{2}\right)\leq M$;
\\
(ii)\;$E\|N^{-1/2}\sum_{k=1}^{N}\lambda_{i}e_{it}\|^{4}\leq M$.
\end{assum}
\begin{assum}\label{invariance} For $\epsilon \in (0,1)$,
$\max_{[\epsilon T]\le k\le T}\left\Vert \frac{1}{\sqrt{NT}}\sum_{t=1}^{k}\sum_{i=1}^{N}[1,f_{t}']'e_{it}\lambda_{i}^{\prime}\right\Vert  =O_{p}(1)$.
 \end{assum}
\begin{assum}\label{limit_distribution2} Let $\eta_{t}=\Sigma_{G}^{-1/2}g_{t}$. Under the null (i.e., $B=C$), $T^{-1/2}\sum\limits _{t=1}^{[\pi T]}(\eta_{t}\eta_{t}^{\prime}-I_{r})\Rightarrow\zeta(\pi)$,
where $\zeta(\pi)$ is an $r\times r$ Gaussian process. \end{assum}

\noindent Assumptions \ref{factors} and \ref{Factor_Loadings} are
standard in the literature on approximate factor models (e.g., Bai,
2003). The time-invariant second moment of $f_{t}$ is commonly used
for identification purposes because a change in the variance of $f_{t}$
is observationally equivalent to a rotation of the factor loadings. A variance change in $f_{t}$ can be absorbed by the matrix $B$
or $C$ in (\ref{DBH}). The uniform convergence of $T^{-1}\sum_{t=1}^{[\pi T]}f_{t}f_{t}^{\prime}$
in Assumption \ref{factors} is similar to Assumption A11 of \cite{qu2007}. Assumptions \ref{Depen_and_Hetero} allows weakly correlated
and heteroskedastic idiosyncratic errors. Assumptions \ref{Weak_Dependence}
and \ref{eigenvalues} correspond to Assumptions D and G of Bai (2003).
Assumption \ref{factor_error}(i) corresponds to Bai's (2003) Assumption F1. Assumption \ref{factor_error}(ii) strengthens Bai's (2003) Assumption F3 and
ensures the consistency of the heteroskedasticity and autocorrelation
consistent (HAC) covariance estimator of $\hat{g}_{t}^{\prime}\hat{g}_{t}-I_{r}$.
Assumption \ref{invariance} follows from Assumption 10 of DBH (2022) or Assumption 8 of HI (2015). Assumption \ref{limit_distribution2}
states that a basic functional central limit theorem holds for the
sums of $g_{t}$ under the null hypothesis of no break.

\subsection{Limiting distribution of sup-LR under the null hypothesis}

It is well known that $\hat{G}$ is an estimator of $GH$, where $H=(\Lambda^{'}\Lambda/N)(G^{'}\hat{G}/T)V_{NT}^{-1}$
and $V_{NT}$ denotes the eigenvalues of $XX'/NT$. Bai's (2003) Proposition
1 shows that $G^{'}\hat{G}/T\to_{p}\Sigma_{\Lambda}^{-1/2}\Upsilon V^{1/2}$,
where $V$ is the probability limit of $V_{NT}$ and an $r\times r$
diagonal matrix of the eigenvalues of $\Sigma_{\Lambda}^{1/2}\Sigma_{G}\Sigma_{\Lambda}^{1/2}$,
and $\Upsilon$ is the eigenvector of $\Sigma_{\Lambda}^{1/2}\Sigma_{G}\Sigma_{\Lambda}^{1/2}$.
Thus, \[
H_{0}\equiv\mathrm{plim}_{N,T\to\infty}H=\Sigma_{\Lambda}^{1/2}\Upsilon V^{-1/2},\]
which implies that \begin{equation}
E(H_{0}^{\prime}g_{t}g_{t}^{\prime}H_{0})=V^{-1/2}\Upsilon^{'}\Sigma_{\Lambda}^{1/2}\Sigma_{G}\Sigma_{\Lambda}^{1/2}\Upsilon V^{-1/2}=V^{-1/2}VV^{-1/2}=I_{r}.\label{eq:HggH=I}\end{equation}
The following theorem establishes the asymptotic distribution of the
sup-LR statistic under the null hypothesis.

\begin{theorem} Under the null hypothesis of no break, if Assumptions
\ref{factors}--\ref{limit_distribution2} hold and $\sqrt{T}/N\to0$
as $N,T\to\infty$, then

(i) \begin{equation}
\sup\limits _{[\epsilon T]\le k\le[(1-\epsilon)T]}LR(k)\Rightarrow\frac{1}{2}\sup\limits _{\pi\in[\epsilon,1-\epsilon]}\frac{[\mathbb{W}(\pi)-\pi\mathbb{W}(1)]'\Omega[\mathbb{W}(\pi)-\pi\mathbb{W}(1)]}{\pi(1-\pi)},\label{eq:distribution}\end{equation}
where  \[
\Omega=\mathrm{plim}_{T\to\infty} \var\left[\frac{1}{\sqrt{T}}\sum_{t=1}^{T}\vec(\eta_{t}\eta_{t}^{\prime}-I_{r})\right]\]
 and $\mathbb{W}(\pi)$ is an $r^{2}$ vector of independent Brownian
motion.

(ii) Define the HAC estimator \begin{equation}
\hat{\Omega}=\hat{\Gamma}_{0}+\sum_{j=1}^{T-1}k\left(\frac{j}{S_{T}}\right)(\hat{\Gamma}_{j}+\hat{\Gamma}_{j}^{\prime}),\ \mathrm{with}\ \hat{\Gamma}_{j}=\frac{1}{T}\sum_{t=j+1}^{T}\vec(\hat{g}_{t}\hat{g}_{t}^{\prime}-I_{r})\vec(\hat{g}_{t-j}\hat{g}_{t-j}^{\prime}-I_{r})^{\prime},\label{eq:hac}\end{equation}
where $k(\cdot)$ is a kernel function and $S_{T}$ is a bandwidth parameter.
Then, \[
\frac{[\mathbb{W}(\pi)-\pi\mathbb{W}(1)]'\hat{\Omega}[\mathbb{W}(\pi)-\pi\mathbb{W}(1)]}{\pi(1-\pi)}\Rightarrow\frac{[\mathbb{W}(\pi)-\pi\mathbb{W}(1)]'\Omega[\mathbb{W}(\pi)-\pi\mathbb{W}(1)]}{\pi(1-\pi)}.\]

\label{distribution_theorem}

\end{theorem}

Note that the limiting null distribution of sup-LR depends on the
long-run singular variance $\Omega$. Part (ii) of Theorem \ref{distribution_theorem}
states that the infeasible $\Omega$ can be replaced with the HAC
estimator in (\ref{eq:hac}) computed using the estimated factors
$\hat{g}_{t}$. Note that $\hat{\Omega}$ is only consistent for $\mathrm{plim}_{T\to\infty}\var[T^{-1/2}\sum_{t=1}^{T}\vec(H_{0}^{\prime}g_{t}g_{t}^{\prime}H_{0}-I_{r})]$
(Theorem 2 of HI (2015)), but inconsistent for $\Omega$.
However, it can be shown that $H_{0}^{\prime}g_{t}=R^{\prime}\eta_{t}$
for some orthonormal matrix $R$. Because a pre-multiplication by $R$
does not change the distribution of an independent standard normal
vector, we can still replace $\Omega$ with $\hat{\Omega}$ when simulating
the limiting null distribution.

Note that the presence of $\Omega$ in (\ref{eq:distribution}) is
due to the potentially misspecified likelihood function (\ref{eq:Likelihood}),
which assumes that $g_{t},t=1,\cdots,T$ are i.i.d. Gaussian under the null hypothesis.
If (\ref{eq:Likelihood}) correctly specifies the likelihood, then
the limiting distribution reduces to \begin{equation}
\sup\limits _{[\epsilon T]\le k\le[(1-\epsilon)T]}LR(k)\Rightarrow\sup\limits _{\pi\in[\epsilon,1-\epsilon]}\frac{[\mathbb{W}_{1}(\pi)-\pi\mathbb{W}_{1}(1)]'[\mathbb{W}_{1}(\pi)-\pi\mathbb{W}_{1}(1)]}{\pi(1-\pi)}\label{eq:andrew limit}\end{equation}
where $\mathbb{W}_{1}(\pi)$ is an $r(r+1)/2$ vector of independent
Brownian motions.
The distribution in (\ref{eq:andrew limit}) is
the same as that used in conventional supreme type tests for a structural
break, and the critical values can be found in \citet{Andrews1993}.

If $g_{t},t=1,\cdots,T$ are i.i.d. Gaussian and the null hypothesis is
true, the sup-LR statistic is numerically close to HI's
(2015) sup-Wald and sup-LM statistics, which compare the subsample
variances of $\hat{g}_{t}$. For example, HI's (2015) sup-Wald
statistic is defined as \[
\sup\limits _{[\epsilon T]\le k\le[(1-\epsilon)T]}W(k)=\sup\limits _{[\epsilon T]\le k\le[(1-\epsilon)T]}A(k)\hat{\Sigma}_{A}^{-1}A(k),\]
where \begin{align*}
A(k) & =\sqrt{T}\left(\frac{1}{k}\sum_{t=1}^{k}\hat{g}_{t}\hat{g}_{t}'-\frac{1}{T-k}\sum_{t=k+1}^{T}\hat{g}_{t}\hat{g}_{t}'\right)\end{align*}
and $\hat{\Sigma}_{A}$ is an unconstrained estimator for the variance
of $A(k)$. Under the null hypothesis, Theorem \ref{thm2}
provides an alternative formulation to approximate the proposed sup-LR
statistic and establishes its connection to the sup-Wald statistic
of HI (2015).

\begin{theorem} \label{thm2} Under the null hypothesis of no break,
if Assumptions \ref{factors}--\ref{limit_distribution2} hold and
$\sqrt{T}/N\to0$ as $N,T\to\infty$, then

\noindent (i)\[
\sup\limits _{[\epsilon T]\le k\le[(1-\epsilon)T]}LR(k)=\sup\limits _{[\epsilon T]\le k\le[(1-\epsilon)T]}\frac{k(T-k)}{2T^{2}}\tr\left(A(k)A(k)\right)+o_{p}(1);\]
(ii) in addition, if $g_{t}$'s are i.i.d. Gaussian, then \[
\sup\limits _{[\epsilon T]\le k\le[(1-\epsilon)T]}LR(k)-\sup\limits _{[\epsilon T]\le k\le[(1-\epsilon)T]}W(k)=o_{p}(1).\]

\end{theorem}

Under the null hypothesis, Theorem \ref{thm2}(i)
provides an approximation of the sup-LR statistic, which involves
a comparison between $\hat{\Sigma}_{1}(k)$ and $\hat{\Sigma}_{2}(k)$.
Part (ii) of Theorem \ref{thm2} shows that the sup-LR test is asymptotically
equivalent to the sup-Wald test under Gaussianity and the null hypothesis of
no break. Although the LR test is asymptotically equivalent to the Wald test under the null, as we show below, the LR test is more powerful under the alternative hypothesis. In addition, the LR test has better finite-sample size properties in simulation.

\subsection{Power against the alternative hypothesis}

To show that the new test is consistent under the alternative
hypothesis, we need the additional assumption set forth below.

\noindent \begin{assum} \label{eig bound}

With probability approaching one (w.p.a.1), the following inequality
holds: \[
\rho_{1}\left(\frac{1}{NT}\sum_{t=1}^{T}\Lambda^{\prime}e_{t}e_{t}^{\prime}\Lambda\right)\le d<\infty,\]
as $N,T\to\infty$, where $d$ is a constant, and $\rho_1(\cdot)$ denotes the largest eigenvalue of a symmetric matrix.

\end{assum}

This assumption is made in DBH (2022). It is a reasonable assumption, because
$\frac{1}{NT}\sum_{t=1}^{T}\Lambda^{\prime}e_{t}e_{t}^{\prime}\Lambda =\frac 1 T \sum_{t=1}^T \left[(\frac 1 {\sqrt{N}} \sum_{i=1}^N \lambda_i e_{it}) (\frac 1 {\sqrt{N}} \sum_{i=1}^N \lambda_i e_{it})'\right]=O_p(1)$.
This ensures
that the smallest eigenvalues of $\hat{\Sigma}_{1}(k)$ and $\hat{\Sigma}_{2}(k)$
are less than $c/T$ for a positive constant $c$ w.p.a.1, according to DBH's (2022) Proposition
1. The theorem set forth below establishes the asymptotic
property of the new test under the alternative hypothesis.

\begin{theorem} \label{thm power} Under Assumptions \ref{factors}--\ref{invariance},

(i) if Assumption \ref{eig bound} holds and the factor loading matrix
has a break such that either $B$ or $C$ in (\ref{DBH}) is singular
and $N\propto T$, then there exists a constant $\underbar{d}>0$
such that \[
\mathrm{Prob}\left(\frac{1}{T\log T}\sup\limits _{[\epsilon T]\le k\le[(1-\epsilon)T]}LR(k)\ge\underbar{d}\right)\to1\]
as $N,T\to\infty$; and

(ii) if the factor loading matrix has a break such that both $B$
and $C$ in (\ref{DBH}) are nonsingular with $\|B\Sigma_{F}B^{\prime}-C\Sigma_{F}C^{\prime}\|>0$,
then there exists a constant $\underbar{d}>0$ such that\[
\mathrm{Prob}\left(\frac{1}{T}\sup\limits _{[\epsilon T]\le k\le[(1-\epsilon)T]}LR(k)\ge\underbar{d}\right)\to1\]
as $N,T\to\infty$.

\end{theorem}

Theorem \ref{thm power} proves that the sup-LR test is consistent
under the alternative hypothesis. Part (i) states that the divergence
rate of LR is $T\log T$, which is faster than the regular rate $T$
of conventional tests. This faster rate is caused by the singularity
of $B$ or $C$. If $B$ is singular, for instance, $\hat{\Sigma}_{1}(k)$
converges in probability to the singular matrix $B\Sigma_{F}B^{\prime}$
for $k\le T_{1}$. In finite samples, $\hat{\Sigma}_{1}(k)$ is nearly
singular, and its smallest eigenvalue has an upper bound $\bar{c}/T$
for some $\bar{c}>0$ under the assumption $N,T\to\infty$ at the
same rate (see the technical details in the appendix). Thus, the term
$-k\log|\hat{\Sigma}_{1}(k)|$ diverges at the faster rate $T\log T$
for $[\epsilon T]\le k\le T_{1}$ with $\epsilon\in(0,0.5)$. In contrast,
conventional tests (such as HI's sup-Wald test) that compare $\hat{\Sigma}_{1}(k)$
and $\hat{\Sigma}_{2}(k)$ do not take advantage of the near singularity
of the sample covariance matrices and thus are less powerful than
our sup-LR test.

\begin{remark}
Define two break point estimators for $T_{1}$ based on our
LR test and HI's Wald test. \begin{align*}
\hat{T}_{1} & =\arg\max_{[\epsilon T]\le k\le[(1-\epsilon)T]}LR(k)\\
\tilde{T}_{1} & =\arg\max_{[\epsilon T]\le k\le[(1-\epsilon)T]}W(k).\end{align*}
$\hat{T}_{1}$ is the QML estimator proposed by DBH (2022),
whereas $\tilde{T}_{1}$ can be viewed as an LS estimator analogous to
the estimator developed by Baltagi et al. (2017). The LS estimator is inconsistent
and has an estimation error $\tilde{T}_{1}-T_{1}=O_{p}(1)$, whereas
the QML estimator is consistent in the sense that $\mathrm{Prob}(\hat{T}_{1}-T_{1}=0)\to1$
as $N,T\to\infty$ if $B$ or $C$ is singular. The ability of the QML estimator to precisely identify the break point is translated into a more powerful sup-LR test.

\end{remark}

\begin{remark}
The alternative hypothesis requires a significant proportion of the cross sections to have a break in their loadings at a common
time, and this setup is frequently utilized in the literature (e.g., Han and Inoue, 2015; Baltagi et al. 2021). If only a small number of series have breaks, there will be a power loss. The reason is the following. In a factor model with
changes in the factor loading matrix, we can express it as a misspecified model with constant loadings, and the changes in the
loadings can be absorbed using the idiosyncratic errors, resulting in correlations in the idiosyncratic errors. However, changes
in a fixed number of loading coefficients only produce a limited level of correlation in the idiosyncratic errors of the
misspecified model. Therefore, the factors can still be estimated consistently up to a rotation, and the sample factor variance
will remain asymptotically the same before and after the break, leading to a loss of power in the test.
\end{remark}

\section{Extensions }

\subsection{Allowing mean change in $g_{t}$}

First we note that testing a change in variance also has power for changes in the mean. Nevertheless, it might be interesting to explicitly allow for a shift in the mean of $g_t$.
 It is common practice to first demean the data and then extract the
principal components so that the full sample average $T^{-1}\sum_{t=1}^{T}\hat{g}_{t}$
is always exactly zero by construction. This property is used here.
To incorporate a mean change in $g_{t}$, we simply redefine the pre-break and post-break variance estimators:
 \begin{eqnarray*}
\tilde{\Sigma}_{1}(k) & = & \frac{1}{k}\sum\limits _{t=1}^{k}(\hat{g}_{t}-\bar{\hat{g}}_{1})(\hat{g}_{t}-\bar{\hat{g}}_{1})',\ \ \mathrm{with\ }\;\bar{\hat{g}}_{1}=\frac{1}{k}\sum\limits _{t=1}^{k}\hat{g}_{t},\\
\tilde{\Sigma}_{2}(k) & = & \frac{1}{T-k}\sum\limits _{t=k+1}^{T}(\hat{g}_{t}-\bar{\hat{g}}_{2})(\hat{g}_{t}-\bar{\hat{g}}_{2})'\;\ \mathrm{with\ }\;\bar{\hat{g}}_{2}=\frac{1}{T-k}\sum\limits _{t=k+1}^{T}\hat{g}_{t},\end{eqnarray*}
where the dependence of $\bar{\hat{g}}_{1}$ and $ \bar{\hat{g}}_{2}$ on $k$ is omitted for notational simplicity.
The likelihood function takes the same form as in (\ref{eq:Likelihood})
 \[
\tilde{\mathcal{L}}(k)=k\log(|\tilde{\Sigma}_{1}(k)|)+(T-k)\log(|\tilde{\Sigma}_{2}(k)|).\]
 Define the LR test that allows for a structural break
in both the mean and the variance of $g_{t}$ as \begin{equation}
\sup-LR_{m}\equiv\sup\limits _{[\epsilon T]\le k\le[(1-\epsilon)T]}LR_{m}(k)=\sup\limits _{[\epsilon T]\le k\le[(1-\epsilon)T]}-k\log(|\tilde{\Sigma}_{1}(k)|)-(T-k)\log(|\tilde{\Sigma}_{2}(k)|).\label{eq:LRm}\end{equation}

We make the following assumption about the weak convergence of the
joint processes.

\begin{assum} \label{joint process} Let $\eta_{t}=\Sigma_G^{-1/2}g_{t}$.
Under the null (i.e., $B=C$),

\[
\frac{1}{\sqrt{T}}\left[\begin{array}{c}
\sum_{t=1}^{[\pi T]}\eta_{t}\\
\vec\sum_{t=1}^{[\pi T]}(\eta_{t}\eta_{t}'-I_{r})\end{array}\right]\Rightarrow\left[\begin{array}{c}
\mathcal{V}(\pi)\\
\zeta(\pi)\end{array}\right],\]
where the right-hand side is a Gaussian process with independent increments.

\end{assum}

\begin{theorem} \label{thm mean break}

Suppose that Assumptions \ref{factors}--\ref{invariance} and \ref{joint process} hold
and $\sqrt{T}/N\to0$ as $N,T\to\infty$ under the null hypothesis
of no break.

\noindent (i) \[
\sup\limits _{[\epsilon T]\le k\le[(1-\epsilon)T]}LR_{m}(k)\Rightarrow\sup\limits _{\pi\in[\epsilon,1-\epsilon]}\frac{[\mathbb{U}(\pi)-\pi\mathbb{U}(1)]'\bar{\Omega}[\mathbb{U}(\pi)-\pi\mathbb{U}(1)]}{\pi(1-\pi)},\]
where $\mathbb{U}$ is an $(r+r^{2})$-dimensional independent Brownian
motion process, and \[
\bar{\Omega}=\mathrm{plim}_{T\to\infty}\var\left(\frac{1}{\sqrt{T}}\left[\begin{array}{c}
\sum_{t=1}^{T}\eta_{t}\\
\vec\sum_{t=1}^{T}(\eta_{t}\eta_{t}'-I_{r})/\sqrt{2}\end{array}\right]\right).\]
(ii) If $\{g_{t}\}$ is uncorrelated with $\{g_{t}g_{t}^{\prime}-\Sigma_{G}\}$,
then \begin{eqnarray*}
\sup\limits _{[\epsilon T]\le k\le[(1-\epsilon)T]}LR_{m}(k) & \Rightarrow & \sup\limits _{\pi\in[\epsilon,1-\epsilon]}\frac{[\mathbb{W}_{m}(\pi)-\pi\mathbb{W}_{m}(1)]'\Omega_{m}[\mathbb{W}_{m}(\pi)-\pi\mathbb{W}_{m}(1)]}{\pi(1-\pi)}\\
 &  & +\frac{1}{2}\frac{[\mathbb{W}(\pi)-\pi\mathbb{W}(1)]'\Omega[\mathbb{W}(\pi)-\pi\mathbb{W}(1)]}{\pi(1-\pi)},\end{eqnarray*}
where $\mathbb{W}_{m}$ and $\mathbb{W}$ are $r$-dimensional and
$r^{2}$-dimensional independent Brownian motion processes, respectively,
and\[
\Omega_{m}=\mathrm{plim}_{T\to\infty}\var\left(T^{-1/2}\sum_{t=1}^{T}\eta_{t}\right),\ \ \Omega=\mathrm{plim}_{T\to\infty}\var\left(T^{-1/2}\sum_{t=1}^{T}\vec(\eta_{t}\eta_{t}^{\prime}-I_{r})\right).\]

\end{theorem}

Part (ii) of Theorem \ref{thm mean break} makes the additional assumption
that $\{g_{t}\}$ and $\{g_{t}g_{t}^{\prime}-\Sigma_{G}\}$ are uncorrelated.
In this case, the limiting null distribution is similar to the result obtained by Qu and Perron (2007), which can be decomposed as
the sum of two components, one corresponding to the mean change and
the other corresponding to the variance change. Part (i) allows a
more general setup with correlated $\{g_{t}\}$ and $\{g_{t}g_{t}^{\prime}-\Sigma_{G}\}$,
and the critical values can be obtained by simulating $r^{2}+r$
dimensional Brownian motions and estimating the long-run variance $\bar{\Omega}$. This long-run variance
can be estimated by replacing $\eta_t$ with $\hat g_t$.

\begin{remark}
If the data are not demeaned before the PC estimation,
then the statistic in (\ref{eq:LRm}) should be defined as \[
\sup-LR_{m}^{nd}\equiv\sup\limits _{[\epsilon T]\le k\le[(1-\epsilon)T]}T\log(|\tilde{\Sigma}|)-k\log(|\tilde{\Sigma}_{1}(k)|)-(T-k)\log(|\tilde{\Sigma}_{2}(k)|),\]
where \[
\tilde{\Sigma}=\frac{1}{T}\sum\limits _{t=1}^{T}(\hat{g}_{t}-\bar{\hat{g}})(\hat{g}_{t}-\bar{\hat{g}})^{\prime},\ \ \mathrm{with\ }\;\bar{\hat{g}}=\frac{1}{T}\sum\limits _{t=1}^{T}\hat{g}_{t}.\]
The asymptotic distribution of $\sup-LR_{m}^{nd}$ under the null hypothesis
of no break is the same as that provided in Theorem \ref{thm mean break}.
\end{remark}
\begin{remark}
Note that $\sum_{t=1}^T \hat{g}_t = 0$ and $\sum_{t=1}^T (\hat{g}_t \hat{g}_t' - I_r) = 0$, so the terms $1/
\sqrt{T}\sum_{t=1}^k \hat{g}_t$ and $1/\sqrt{T} \vech\sum_{t=1}^k (\hat{g}_t\hat{g}_t - I_r)$ weakly converge
to Brownian bridge type processes (see the proof of Theorem 4 for the technical details). Thus, a
Wald test that allows a factor mean change can be defined without directly comparing
$\hat{\Sigma}_1(k)$ and $\hat{\Sigma}_2(k)$. Define \begin{eqnarray*}
Wald_m(k)=\frac{T^{2}}{k(T-k)}\left[\begin{array}{cccccccccc}
\frac{1}{\sqrt{T}}\sum_{t=1}^{k}\hat{g}_{t}\\
\frac{1}{\sqrt{T}}\vech\sum_{t=1}^{k}(\hat{g}_{t}\hat{g}_{t}'-I_{r})\end{array}\right]'\hat{\Omega}^{-1}_w\left[\begin{array}{cccccccccc}
\frac{1}{\sqrt{T}}\sum_{t=1}^{k}\hat{g}_{t}\\
\frac{1}{\sqrt{T}}\vech\sum_{t=1}^{k}(\hat{g}_{t}\hat{g}_{t}'-I_{r})\end{array}\right]\end{eqnarray*}
where $\hat{\Omega}_w$ is the corresponding long-run variance. Note that here we use $\vech$ instead of $\vec$. The limit is similar to that of the usual Wald test,
\begin{eqnarray*}
Wald_m(k)\Rightarrow\frac{\|W(\pi)-\pi W(1)\|^{2}}{\pi(1-\pi)}\end{eqnarray*}
 where $W(\pi)$ is an $r+r(r+1)/2=r(r+3)/2$ standard Brownian motion.
This test is nuisance-parameter free.

Consider the simple weighting by $k(T-k)/T^{2}$, let $WWald=k(T-k)/T^{2}\cdot Wald_m(k)$.
Take the maximum over the range say $k=r+1,r+2,\cdots,T-r-1$.
The limiting distribution is simply \[
\sup\limits _{\pi\in[0,1]}\left\Vert W(\pi)-\pi W(1)\right\Vert ^{2},\]
and the critical values do not depend on trimming parameter $\epsilon$.
\end{remark}

\subsection{Testing multiple changes}

In this section, we extend the sup-LR test to multiple
changes. We consider testing the null hypothesis of no change versus
the alternative hypothesis of a prespecified number of changes. To allow
for $m$ changes under the alternative hypothesis, let $0=k_{0}<k_{1}<k_{2}<\cdots<k_{m}<k_{m+1}=T$
and define \[
LR(k_{1},...,k_{m})=-\sum_{j=0}^{m}(k_{j+1}-k_{j})\log(|\hat{\Sigma}_{j+1}|),\]
where \[
\hat{\Sigma}_{j+1}=\frac{1}{k_{j+1}-k_{j}}\sum\limits _{t=k_{j}+1}^{k_{j+1}}\hat{g}_{t}\hat{g}_{t}^{\prime},\ \mathrm{for}\ j=0,...,m.\]
We define the set $\Theta_{k,\epsilon}=\{(k_1,\cdots,k_m):|k_{j+1}-k_j|\geq [T \epsilon],k_m\leq [(1-\epsilon)T]\}$ and $\Theta_{\pi,\epsilon}=\{(\pi_1,\cdots,\pi_m):|\pi_{j+1}-\pi_j|\geq  \epsilon,\pi_m\leq 1-\epsilon\}$.

\begin{theorem}\label{thm multi breaks} Let $0=\pi_{0}<\pi_{1}<...<\pi_{m}<\pi_{m+1}=1$ and
$\pi_j=k_j/T$.
Under Assumptions \ref{factors}--\ref{limit_distribution2}
and $\sqrt{T}/N\to0$ as $N,T\to\infty$, we have
\begin{eqnarray}\label{thm multi breaks_no_correlation}
\sup\limits_{(k_1,\cdots,k_m)\in \Theta_{k,\epsilon}} LR(k_{1},\cdots,k_{m})
 & \Rightarrow &\sup\limits_{(\pi_1,\cdots,\pi_m)\in \Theta_{\pi,\epsilon}}
\sum_{j=0}^{m}\frac{1}{2(\pi_{j+1}-\pi_{j})}  \|  \Omega^{1/2}[\tilde{\mathbb{B}}(\pi_{j+1})-\tilde{\mathbb{B}}(\pi_{j})]   \|^2,
\end{eqnarray}
where $\tilde{\mathbb{B}}(\pi)=\mathbb{W}(\pi)-\pi \mathbb{W}(1)$ is an $r^2$ Brownian
bridge process and $\Omega$ is defined in Theorem \ref{distribution_theorem}.
\end{theorem}

The limiting distribution in Theorem \ref{thm multi breaks}
is different from the case with an observed $g_{t}$, as in Bai and Perron (1998).
Each term in the sum involves an increment of the Brownian bridge instead of Brownian motion because of the PC normalization in (\ref{restrictions}).

To allow for both mean and variance changes, we define the
following statistic

\[
LR_{m}(k_{1},...,k_{m})=-\sum_{j=0}^{m}(k_{j+1}-k_{j})\log(|\tilde{\Sigma}_{j+1}|),\]
where \[
\tilde{\Sigma}_{j+1}=\frac{1}{k_{j+1}-k_{j}}\sum\limits _{t=k_{j}+1}^{k_{j+1}}(\hat{g}_{t}-\bar{\hat{g}}_{j+1})(\hat{g}_{t}-\bar{\hat{g}}_{j+1})^{\prime},\ \ \mathrm{with\ }\;\bar{\hat{g}}_{j+1}=\frac{1}{k_{j+1}-k_{j}}\sum\limits _{t=k_{j}+1}^{k_{j+1}}\hat{g}_{t},\ \mathrm{for}\ j=0,...,m.\]
\begin{theorem}\label{allow_mean_change_distribution_theorem_f_ff_sigma_correlated}
Under Assumptions \ref{factors}--\ref{invariance} and \ref{joint process}, if $0=\pi_{0}<\pi_{1}<...<\pi_{m}<\pi_{m+1}=1$
and $\sqrt{T}/N\to0$ as $N,T\to\infty$, \begin{eqnarray}\label{thm multi breaks_with_correlation}
\sup\limits_{(k_1,\cdots,k_m)\in \Theta_{k,\epsilon}}  LR_{m}(k_{1},...,k_{m}) & \Rightarrow\sup\limits_{(\pi_1,\cdots,\pi_m)\in \Theta_{\pi,\epsilon}} \sum\limits _{j=0}^{m}\frac{\|\bar{\Omega}^{1/2}[\mathbb{B}(\pi_{j+1})-\mathbb{B}(\pi_{j})]\|^{2}}{\pi_{j+1}-\pi_{j}},\end{eqnarray}
where $\mathbb{B}(\pi)=\mathbb{U}(\pi)-\pi\mathbb{U}(1)$ is an $(r+r^2)$ Brownian
bridge process, and $\bar{\Omega}$ is defined in Theorem \ref{thm mean break}.
\end{theorem}

These results depend on $m$, but $m$ does not have to be correctly specified. We can also consider the double max
type of test and the conditional test for multiple breaks as in Bai and Perron (1998), but we leave these as future research topics.

\subsection{Determining the number of breaks}
In practice, the number of breaks in a factor model is often unknown. We follow Bai's (1997) sequential testing procedure to provide a consistent estimate for the true number of breaks. The procedure involves treating the model as if there is only one change point at each time.

To be specific, we first identify the initial break point, $\hat k_1$, using the QML method in DBH.
To determine the presence of any additional breaks, we split the entire sample into two subsamples: $[1,\hat k_1]$ and $[\hat k_1, T]$. For each subsample, we conduct the sup-LR test with $m=1$ as outlined in (\ref{thm multi breaks_no_correlation}) or (\ref{thm multi breaks_with_correlation}), and employ the QML method to estimate a break point for the subsample where the null hypothesis is rejected at a significance level of $\alpha$.
We then split the corresponding subsample into further subsamples at the newly estimated break point and repeat the LR tests for each subsample. This process continues until the LR test is not rejected for all subsamples. The number of break points is equal to the number of subsamples minus one, and the location of the change points can be estimated in the procedure.
The following corollary shows that the number of breaks can be consistently estimated when $\alpha$ converges to zero slowly.
\begin{corollary}\label{determine_number_of_breaks}
Suppose that the size of the test $\alpha$ converges to zero slowly ($\alpha\rightarrow 0$ yet $\lim\inf_{T\rightarrow \infty}T\alpha>0$), then under the assumptions of Theorems \ref{thm multi breaks} and \ref{allow_mean_change_distribution_theorem_f_ff_sigma_correlated},
\begin{align*}
P(\hat m=m_0)\rightarrow 1,\,\,as\,(N,T)\rightarrow \infty.
\end{align*}
\end{corollary}
The proof of this corollary is similar to Proposition 11 in Bai (1997), and the details are omitted here. In practice, one can choose a small $\alpha$ value. In our simulation, we set $\alpha$ to be 0.05.

\section{Monte Carlo simulations}

In this section, we investigate the finite sample properties of the
proposed LR and LR$_m$ tests for the factor models
with a single break point. We compare the performance of our LR and LR$_m$ tests with HI's Wald test and the LM test
for various sample sizes and different setups for the factors and factor loadings in Tables \ref{size_normal_tau030}-\ref{power_mean_shift_change_increase_samples_dd}. As shown below, we use Wald(HAC) and LM(HAC) to denote HI's Wald and LM tests, respectively, using HAC variances with a Bartlett kernel. All of the HAC
estimates are based on Newey and West's (1994) method.

The asymptotic critical values for HI's tests are provided
by Andrews (1993), and the asymptotic critical values for the proposed tests are obtained by simulation. In all of the simulations, we set the time length of the Brownian motions equal to 2,000 and repeat them 1,000 times to obtain the critical values.

We investigate these tests' size in Tables \ref{size_normal_tau030}-\ref{size_t_10_tau015}, and their  power in Tables \ref{power_mean_shift_Normal_increase_samples}-\ref{power_mean_shift_change_increase_samples_dd}. The number of factors is selected by $IC_{p_1}$ of Bai and Ng (2002). In all of the Monte Carlo experiments, we calculate size and power based on 2,000 replications for
each data generating process (DGP).

Each factor is generated by the following
AR(1) process: \begin{eqnarray*}
f_{tp}=\rho f_{t-1,p}+u_{t,p},\quad for\quad t=2,\cdots,T;\quad p=1,\cdots,r_{0},\end{eqnarray*}
 where $u_{t}=(u_{t,1},\cdots,u_{t,r_{0}})^{'}$ is i.i.d.
for $t=2,\cdots,T$ and $f_{1}=(f_{1,1},\cdots,f_{1,r_{0}})^{'}$
is i.i.d. $N(0,I_{r_{0}})$. The scalar $\rho$
captures the serial correlation of factors, and the idiosyncratic
errors $e_{i,t}$ are i.i.d. $N(0,r_0)$.
We set the break date $T_1=T/2$ in all of the setups. We consider the following DGPs for factor loadings and
investigate the performance of the tests for the different types of
breaks discussed in the previous section.

Tables \ref{size_normal_tau030} and \ref{size_t_10_tau030} present the simulation results under the null hypothesis when the permissible break dates are between $0.30T$ and $0.70T$ for different nominal sizes of $10\%$, $5\%$, and $1\%$.
We set $u_t\sim N(0,1-\rho^2)$ and $u_t\sim \sqrt{\frac{1-\rho^2}{1.25}}t(10)$ for Tables \ref{size_normal_tau030} and \ref{size_t_10_tau030}, respectively, and set $\lambda_{i1}\overset{\mathrm{i.i.d.}}{\sim} N(0,I_{r_0})$ for $i=1,\cdots,N$ and $r_0=3$. Under the null hypothesis, $\Lambda_1=\Lambda_2$. From Tables \ref{size_normal_tau030} and \ref{size_t_10_tau030}, we can see that all of these tests have effective sizes approaching the nominal levels as the sample size increases. Tables \ref{size_normal_tau015} and \ref{size_t_10_tau015} show similar simulation results under the null hypothesis when the permissible break dates are from $0.15T$ to $0.85T$ for different nominal sizes.

Table \ref{power_mean_shift_Normal_increase_samples} reports how the power of these tests changes as the sample size increases. We set $\lambda_{i2}=\lambda_{i1}+\varepsilon_i$, $\varepsilon_i \overset{\mathrm{i.i.d.}}{\sim}N(0,I_{r_0})$ for $i=1,\cdots,N$, $r_0=2$, $u_t\sim N(0,1-\rho^2)$ and the permissible break dates are from $0.30T$ to $0.70T$. This setup shows that the post-break factor loadings undergo a random shift.
In each estimation, we use Bai and Ng's (2002) $IC_{p_1}$ to estimate the number of factors.
Table \ref{power_mean_shift_Normal_increase_samples} shows that all of these tests become
powerful when the sample size increases. However, we can find that LR and LR$_m$ are more powerful than Wald(HAC) and LM(HAC) under small sample sizes. When $N=T=100$, the power of Wald(HAC) and LM(HAC) is less than $70\%$, but that of LR test is almost $100\%$. These results consolidate Theorem \ref{thm power}.

Table \ref{power_mean_shift_Normal_increase_break_magnitude} reports how power changes in these tests as the magnitude of the break in factor loadings increases. We set $\lambda_{i2}=\lambda_{i1}+\varepsilon_i$, $\varepsilon_i \overset{\mathrm{i.i.d.}}{\sim}N(0,b\cdot I_{r_0})$ for $i=1,\cdots,N$, $r_0=2$, $u_t\sim N(0,1-\rho^2)$, and the permissible break dates are from $0.15T$ to $0.85T$. When $b=0$, the rejection frequency represents the size under the null hypothesis.
As $b$ increases, all of these tests become powerful, although the power of Wald(HAC) and LM(HAC) is not monotonically increasing in $b$. When $\rho=0$ and $N=T=100$, LR and LR$_m$ are more powerful than Wald(HAC) and LM(HAC). The power of LR and LR$_m$ increases substantially and approaches one as $b$ approaches one. Although the power of Wald(HAC) and LM(HAC) increases as $b$ increases, the power when $b=1$ is smaller than the case of $b=0.8$.
As T increases from $100$ to $200$, the power of Wald(HAC) and LM(HAC) increases, but they are still less powerful than LR and LR$_m$.
When $\rho=0.5$ and $N=T=100$, LR and LR$_m$ also demonstrate superior power relative to the Wald and LM tests.
The power of Wald(HAC) and LM(HAC) is poor when the sample size is small, even if the break magnitude is large.
When $b=1$ and $N=T=100$, the power of Wald(HAC) is less than $55\%$, and the power of LM(HAC) is less than $30\%$. When $T$ increases to $200$, the power of Wald(HAC) and LM(HAC) improves but is still smaller than the power of LR and LR$_m$.

Table \ref{power_rotational_change_increase_aa} reports the power when the loading matrix undergoes a rotational change. We set $\Lambda_2=a\Lambda_1$, $r_0=3$, $u_t\sim N(0,1-\rho^2)$. When $a=1$, the rejection frequency represents the size under the null hypothesis. As the value of \emph{a} increases, all of the tests become more powerful, except for Wald(HAC) and LM(HAC) when $N=T=100$. The LR test has the highest power under this DGP. The power of the LR$_m$ test increases at a slower rate as $a$ increases.
Our Theorem 3 demonstrates that the sup-LR
diverges at the same rate as the conventional sup-Wald when both pre-break and post-break pseudo-factors have nonsingular variances.
However, our simulation results indicate that sup-LR remains more powerful than sup-Wald under this scenario, as reported in Table \ref{power_rotational_change_increase_aa}.
Specifically, when the loading matrix undergoes a rotational change, sup-LR is notably more powerful than sup-Wald with HAC variance.
While we acknowledge that the theoretical power comparison under this setup may require analysis under local alternatives, we leave this as a future research topic.

Table \ref{power_singular_rotational_change_increase_samples} presents the power against changes in the number of factors. The post-break loadings are equal to pre-break loadings multiplied by an $r\times r$ matrix, i.e., $\Lambda_2=\Lambda_1 C$. We set $C=[1,0,0;0,1,0;0,0,0]$, $r_0=3$, $\rho=0.5$.
From Table \ref{power_singular_rotational_change_increase_samples}, we find that the LR and LR$_m$ tests have higher than Wald(HAC) and LM(HAC).

Table \ref{power_mean_shift_change_increase_samples_dd} shows the power when the factors undergo a mean change.
The factors are generated in the same way as above, except that we add a constant to the post-break factors, i.e., $\tilde{F}_1=F_1$ and $\tilde{F}_2=F_2+d$, where $F_1$ and $F_2$ are the original factors and $d$ is a constant. We set $r_0=2$, $d=0.5$, $u_t\sim N(0,1-\rho^2)$. It is remarkable that all of the test statistics have power under this type of break. Unsurprisingly, LR$_m$ is the most powerful statistic under this DGP.

\begin{figure}[H]
\centering
\subfigure[$N=100,T=100$]{
\includegraphics[width=0.5\textwidth,height=0.3\textwidth]{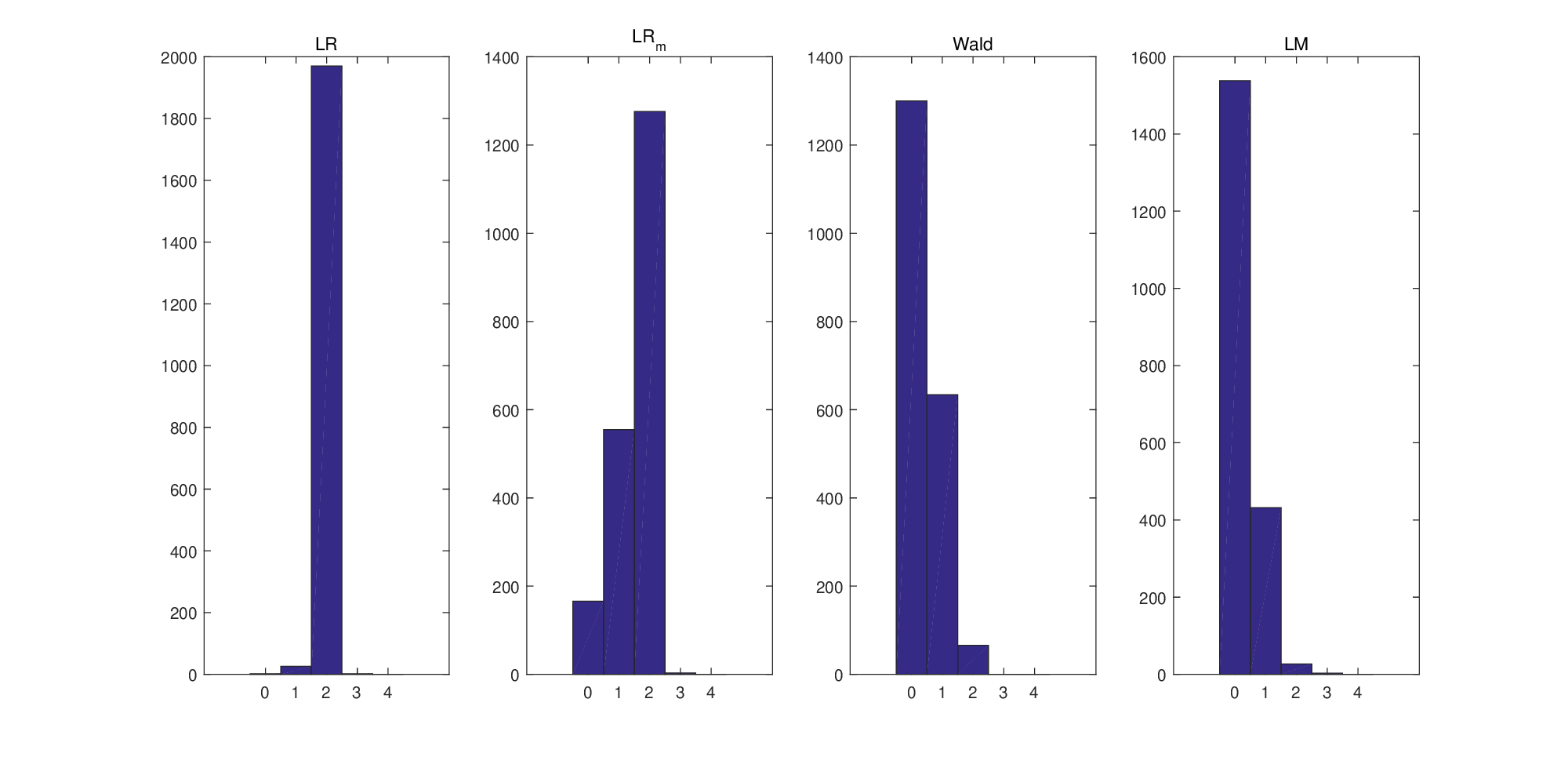}
}\hspace{-12mm}
\quad
\subfigure[$N=200,T=100$]{
\includegraphics[width=0.5\textwidth,height=0.3\textwidth]{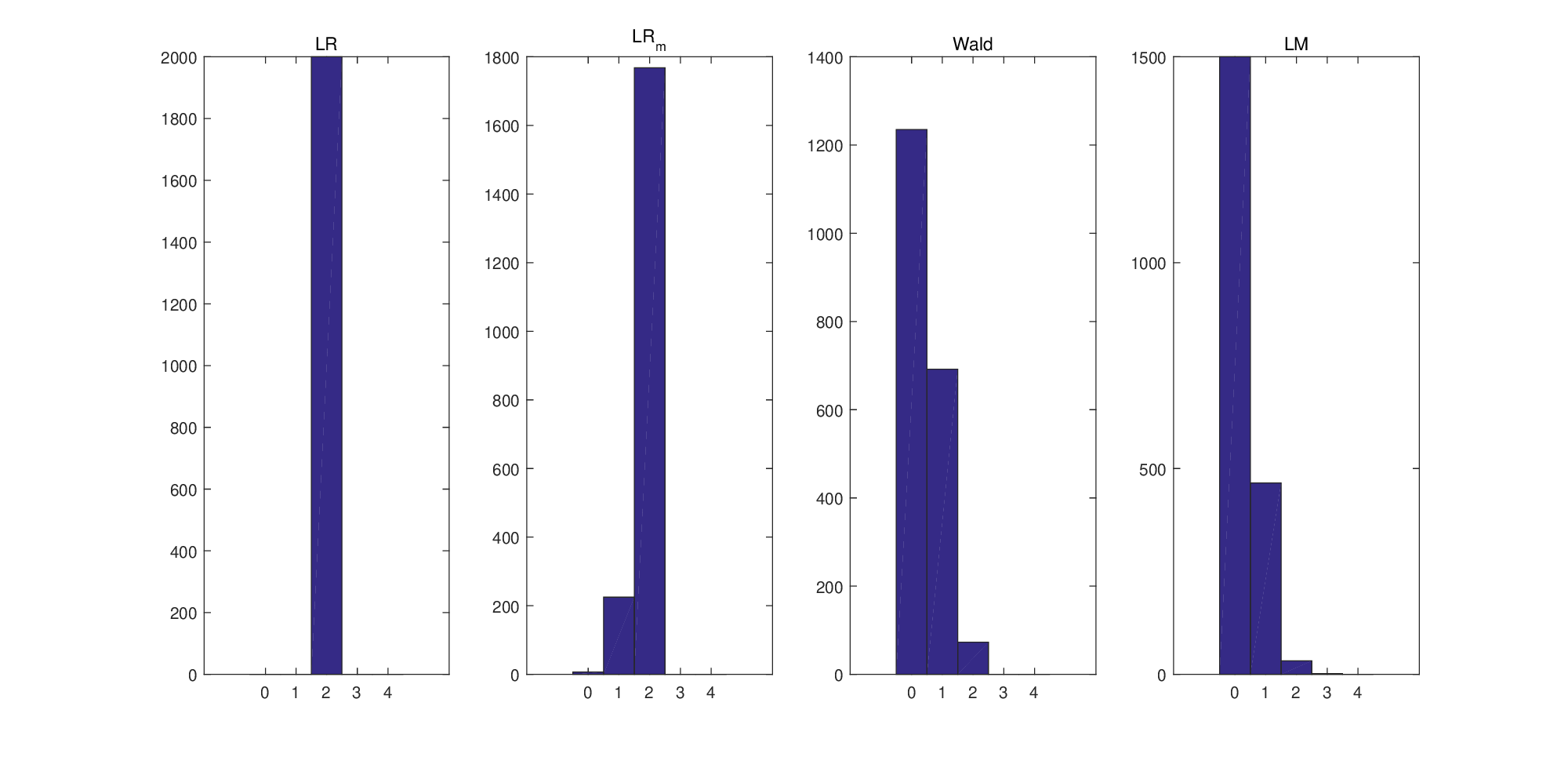}
}\hspace{-12mm}
\quad
\subfigure[$N=100,T=200$]{
\includegraphics[width=0.5\textwidth,height=0.3\textwidth]{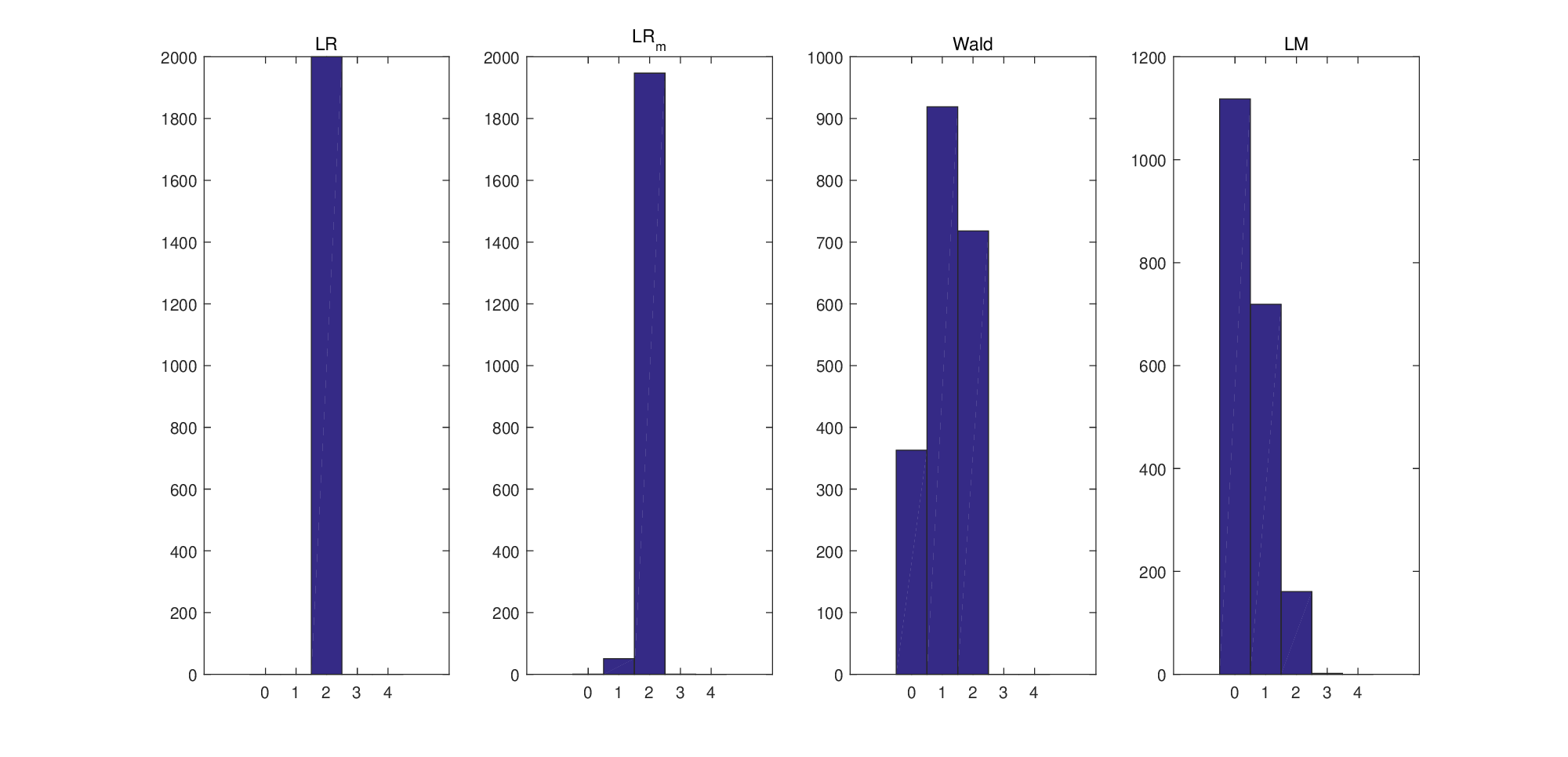}
}\hspace{-12mm}
\quad
\subfigure[$N=200,T=200$]{
\includegraphics[width=0.5\textwidth,height=0.3\textwidth]{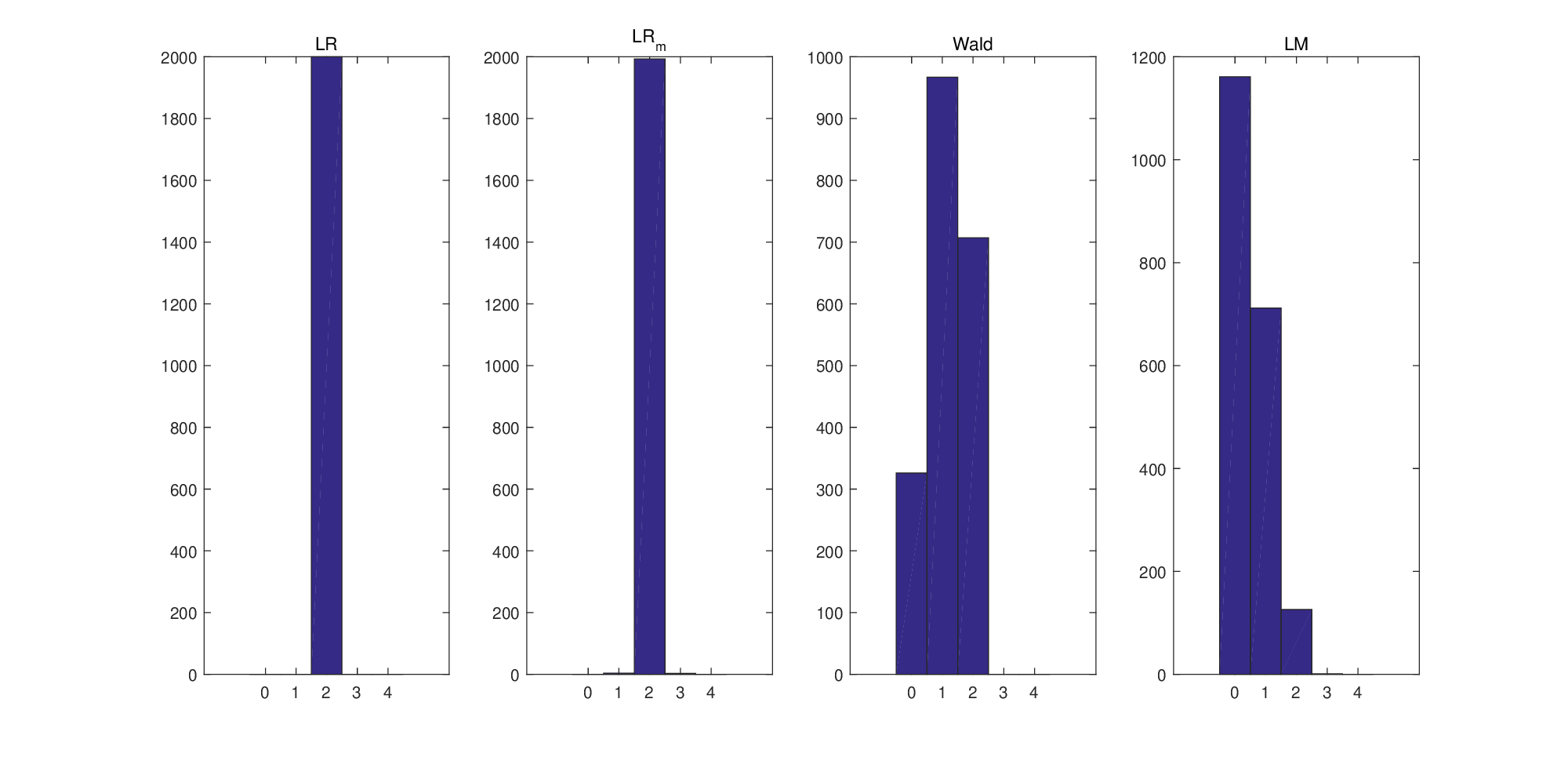}
}\hspace{-12mm}
\quad
\subfigure[$N=100,T=300$]{
\includegraphics[width=0.5\textwidth,height=0.3\textwidth]{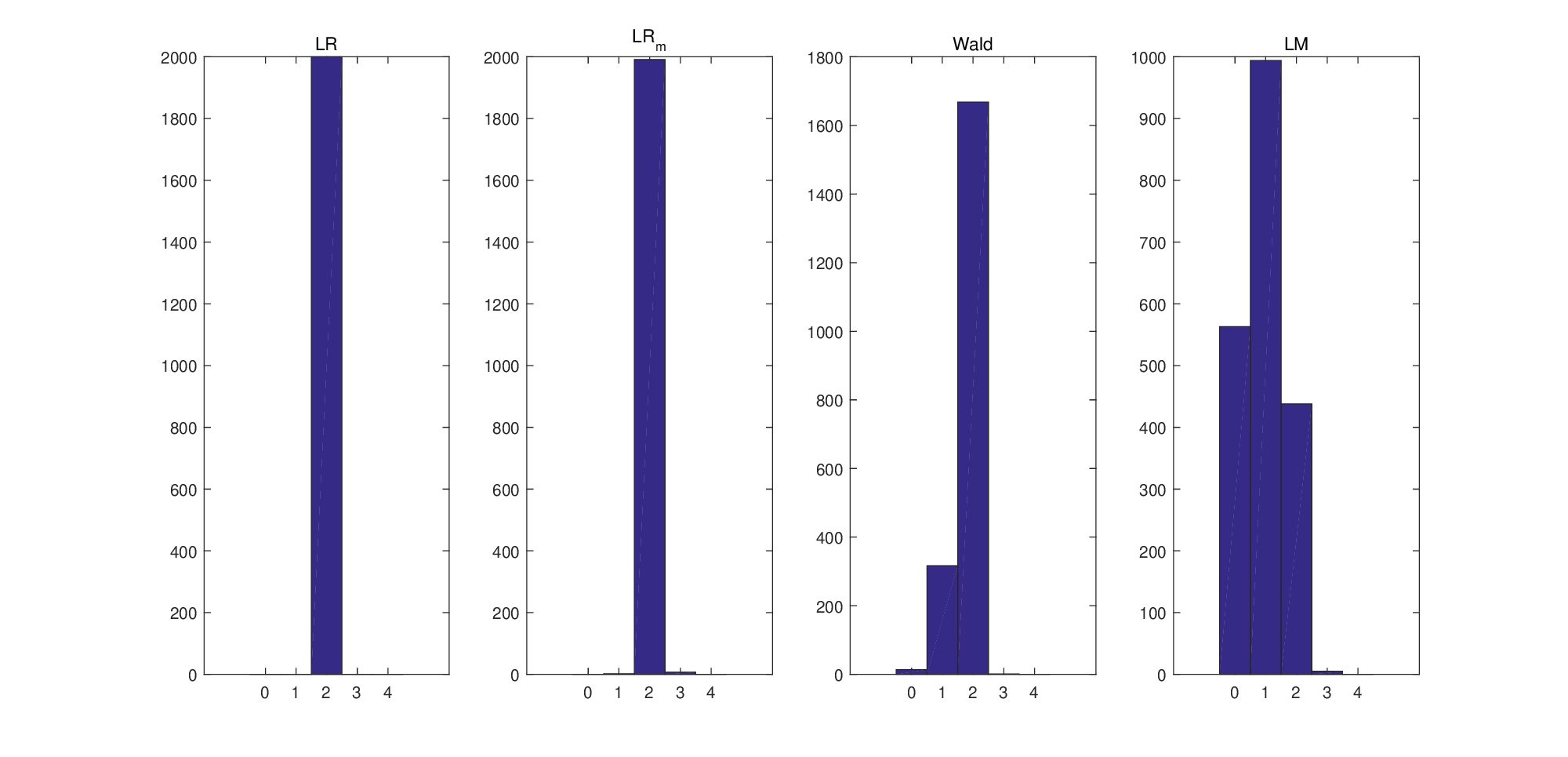}
}\hspace{-12mm}
\quad
\subfigure[$N=200,T=300$]{
\includegraphics[width=0.5\textwidth,height=0.3\textwidth]{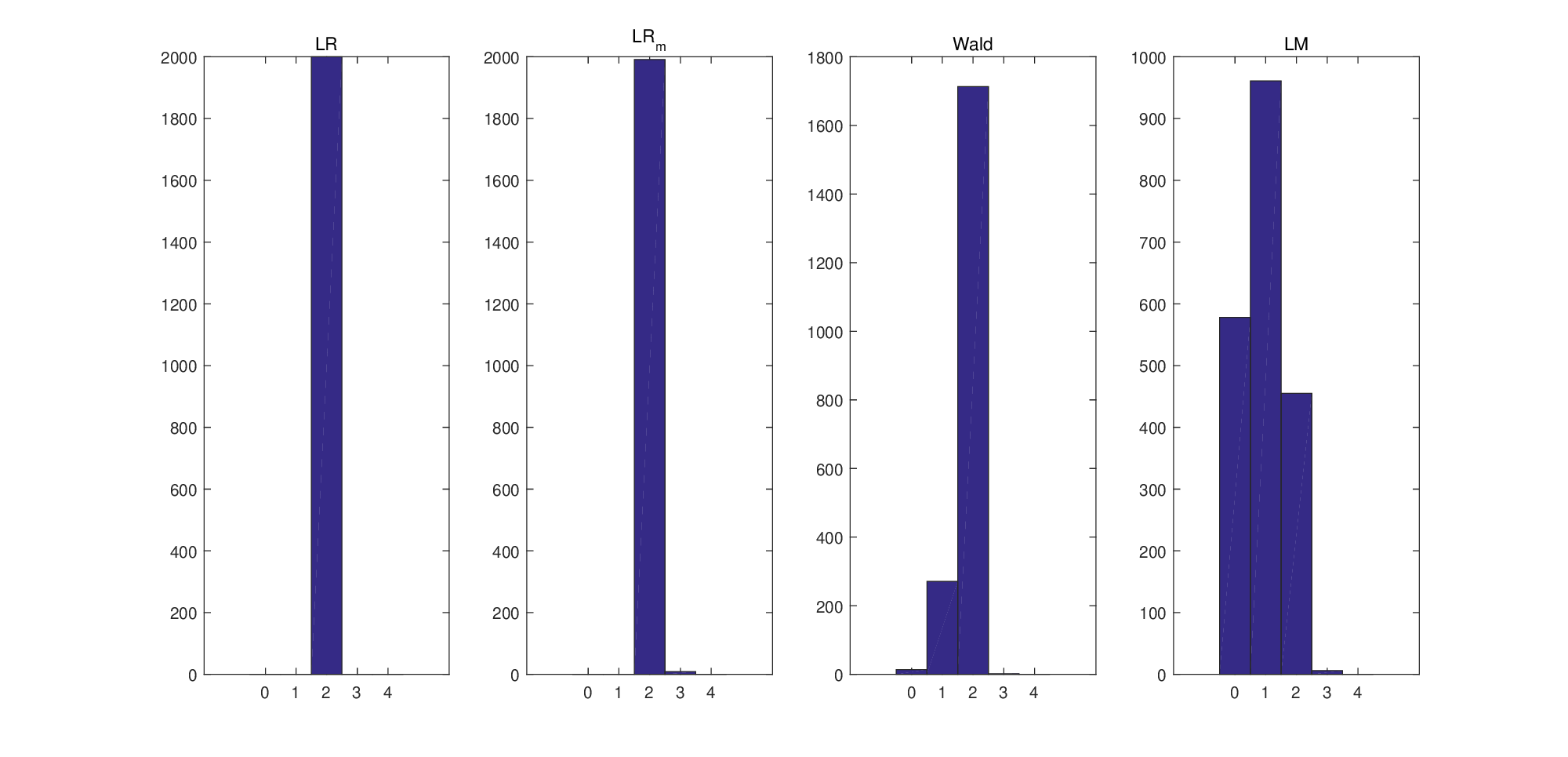}
}\hspace{-12mm}
\quad
\subfigure[$N=100,T=500$]{
\includegraphics[width=0.5\textwidth,height=0.3\textwidth]{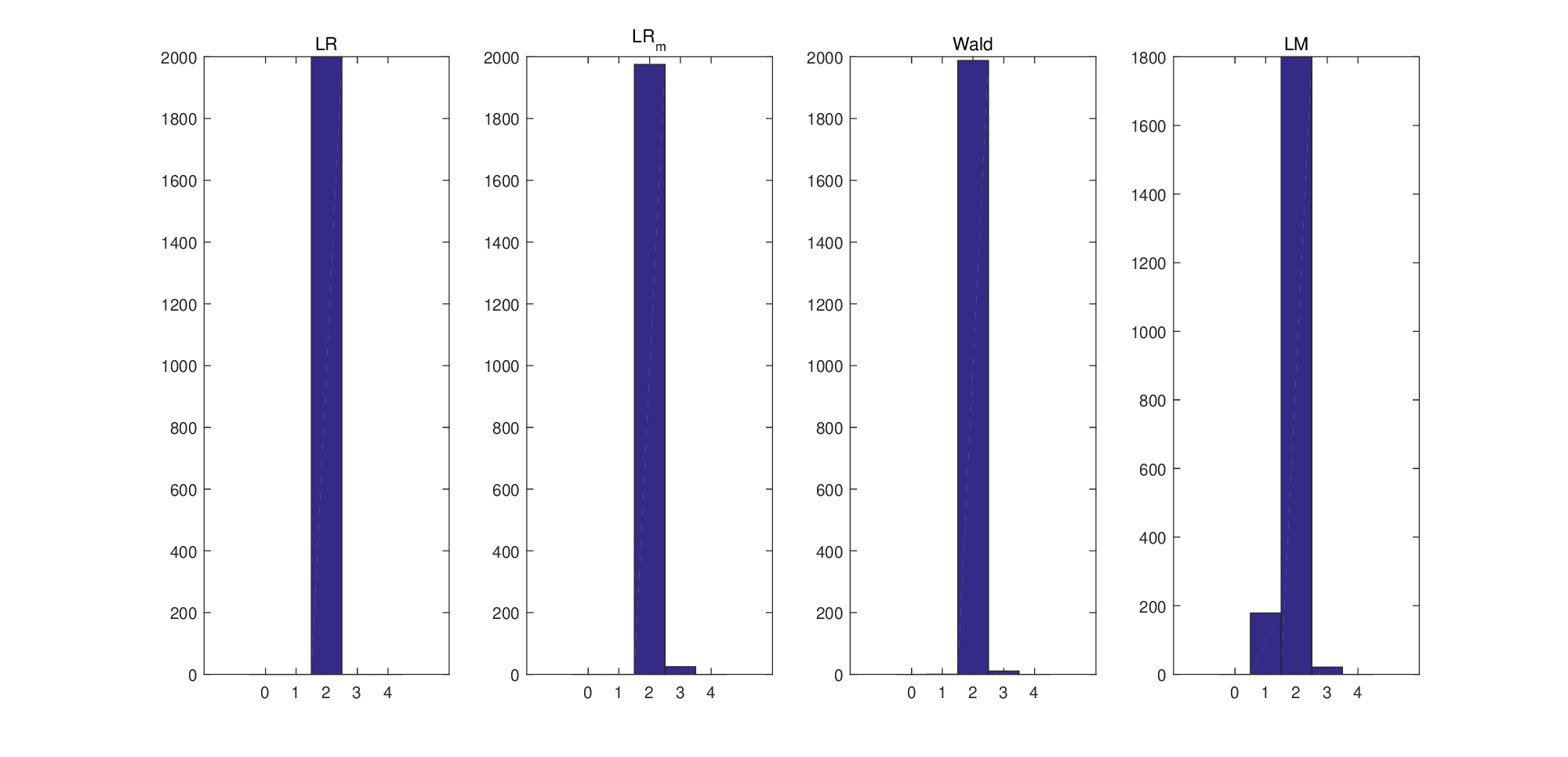}
}\hspace{-12mm}
\subfigure[$N=200,T=500$]{
\includegraphics[width=0.5\textwidth,height=0.3\textwidth]{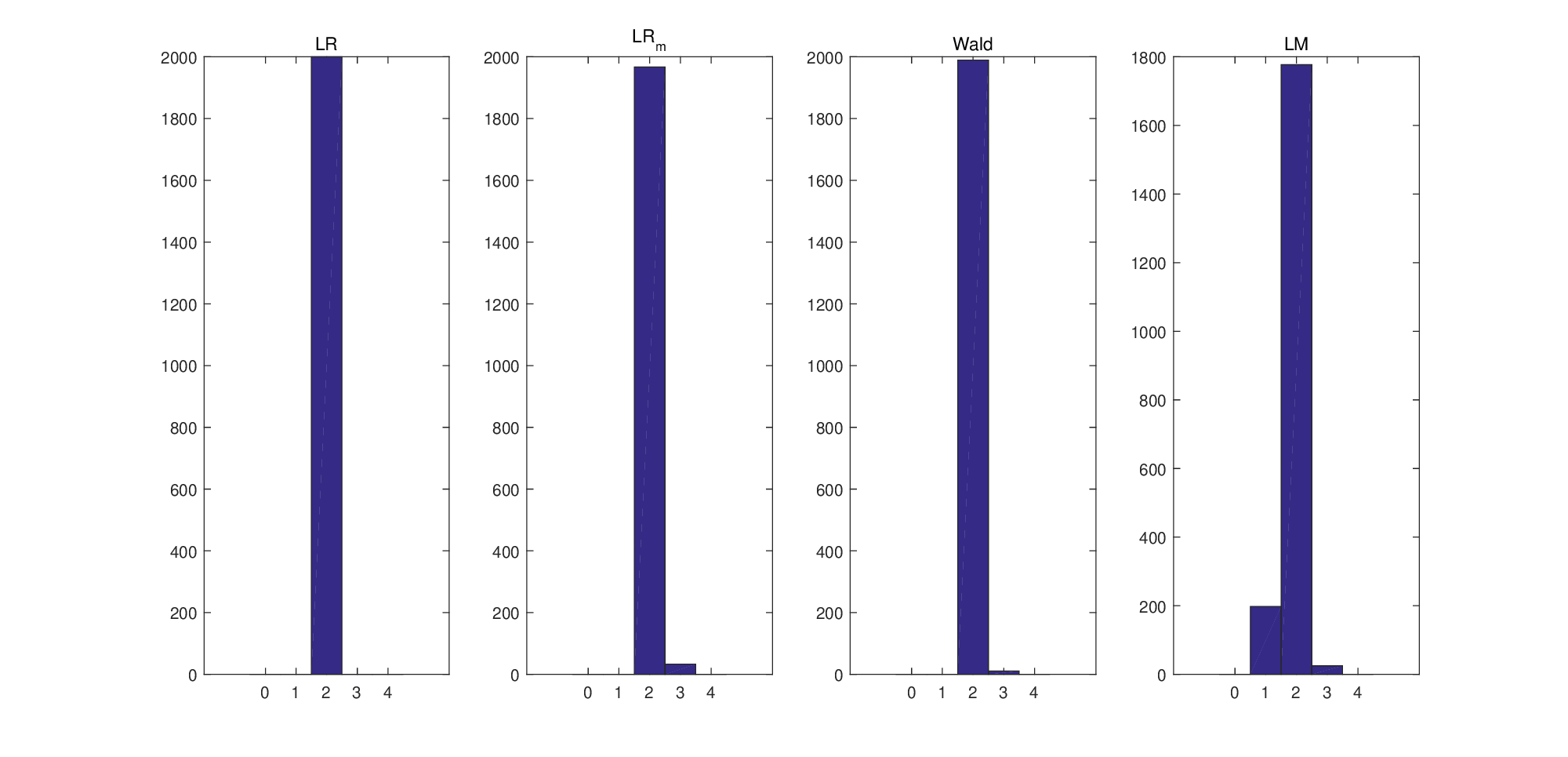}
}
\caption{Histograms of the estimated numbers of breaks.}
\label{multiple_breaksN100}
\end{figure}

Figure \ref{multiple_breaksN100} displays the number of breaks estimates using the sequential
method based on sup-LR, sup-LR$_{m}$, sup-Wald(HAC) and sup-LM(HAC)
tests. The data generating process (DGP) is the same as that used in Table 5, except that the
loadings experience two breaks at ${[}T/3{]}$ and ${[}2T/3{]}$. The loadings in the first regime
are generated as $\lambda_{1i}\sim N(0,I_{r_{0}})$ for $i=1,...,N$. The loadings in the
second and third regimes are generated as $\lambda_{2i}=\lambda_{1i}+\delta_{2i}$
and $\lambda_{3i}=\lambda_{1i}+\delta_{3i}$, respectively, where
both $\delta_{2i}$ and $\delta_{3i}$ are iid draws from $N(0,\ 0.8I_{r_{0}})$. We set $\rho=0.5$. The sequential procedure uses $5\%$ tests, as in Bai (1997).

In Figure \ref{multiple_breaksN100}, the estimated number of breaks becomes more accurate
as $T$ and $N$ increase for all tests. However, the estimates obtained
using LR and LR$_{m}$ are more accurate than those obtained using Wald and LM. The
advantage of LR and LR$_{m}$ is particularly pronounced when the sample size is small.

\begin{table}[H]
\caption{Size of structural break tests with unknown break date in a factor model with Gaussian factors. $\epsilon=0.30$.}
\centering
\label{size_normal_tau030}
\begin{center}
\resizebox{\textwidth}{!}{
\begin{tabular}{l l r r r r r r r r r r r r r r r r r r r r r  r r r} \hline

$N,T$        &  \multicolumn{3}{c}{sup-Wald(HAC)}  &  \multicolumn{3}{c}{sup-LM(HAC)} &  \multicolumn{3}{c}{sup-LR}&   &\multicolumn{3}{c}{sup-LR$_m$}\\

            &    10\% & 5\%  &  1\%  & 10\% &  5\% &  1\%  &  10\%    & 5\%  & 1\% &  10\%    & 5\%  & 1\% & \\\hline

$\rho=0$&       &   &   &  &  &   &     &   &  &    &  &  \\



100,100     &0.012   &0.004   &0.000     &0.011  &0.005  &0.003   &0.049     &0.021   &0.002   &0.128     &0.084   &0.050  \\

100,200    &0.034   &0.015   &0.002     &0.027  &0.010  &0.003   &0.070     &0.029   &0.005    &0.054     &0.023   &0.011 \\

100,500  &0.064   &0.025   &0.003    &0.064  &0.026  &0.002   &0.086     &0.039   &0.004    &0.055     &0.017   &0.003  \\

200,200    &0.035   &0.012   &0.001     &0.024  &0.010  &0.001   &0.063     &0.022   &0.003    &0.035     &0.016   &0.007 \\

200,500   &0.072   &0.028   &0.005    &0.072  &0.026  &0.004   &0.083     &0.044   &0.008    &0.054     &0.018   &0.001 \\

500,500  &0.054   &0.024   &0.003     &0.055  &0.021  &0.003   &0.068     &0.034   &0.005    &0.044     &0.018   &0.002 \\\hline

$\rho=0.5$    &  &   &   &  &  &   &     &   &  &    &  &   \\



100,100 &0.031   &0.010   &0.002     &0.033  &0.011  &0.002   &0.069     &0.029   &0.004    &0.121     &0.083   &0.047  \\

100,200  &0.075   &0.028   &0.002    &0.055  &0.021  &0.004   &0.092     &0.041   &0.005    &0.071     &0.035   &0.010   \\

100,500  &0.130   &0.068   &0.014     &0.109  &0.053 &0.012   &0.112     &0.059   &0.011    &0.078     &0.031   &0.008 \\

200,200  &0.081   &0.035   &0.003     &0.071  &0.031  &0.009   &0.097     &0.041   &0.010    &0.056   &0.027   &0.009   \\

200,500  &0.129   &0.061   &0.014    &0.103  &0.051  &0.009   &0.108     &0.050   &0.014   &0.077     &0.032   &0.005    \\

500,500    &0.123   &0.067   &0.017   &0.100  &0.050  &0.012   &0.114     &0.054   &0.013   &0.076     &0.038   &0.010  \\\hline

\end{tabular}}
\end{center}
\end{table}
\begin{table}[H]
\caption{Size of structural break tests with unknown break date in a factor model with $t$-distributed factors.  $\epsilon=0.30$.}
\centering
\label{size_t_10_tau030}
\begin{center}
\resizebox{\textwidth}{!}{
\begin{tabular}{l l r r r r r r r r r r r r r r r r r r r r r  r r r} \hline

$N,T$      &  \multicolumn{3}{c}{sup-Wald(HAC)}   &  \multicolumn{3}{c}{sup-LM(HAC)} &  \multicolumn{3}{c}{sup-LR} &  \multicolumn{3}{c}{sup-LR$_m$}\\

             & 10\% & 5\%  &  1\%  & 10\% &  5\% &  1\%  &  10\%    & 5\%  & 1\% &  10\%    & 5\%  & 1\%  \\\hline

$\rho=0$&     &   &   &  &  &   &     &   &  &    &  &   \\



100,100   &0.012   &0.001   &0.000     &0.013  &0.004  &0.001   &0.041     &0.012   &0.000   &0.110     &0.073   &0.046  \\

100,200    &0.02   &0.011   &0.000   &0.031  &0.009  &0.001   &0.054     &0.018   &0.001   &0.029     &0.012   &0.006 \\

200,200   &0.026   &0.009   &0.001    &0.019  &0.008  &0.001   &0.057     &0.021   &0.004   &0.036     &0.016   &0.009  \\

200,500  &0.061   &0.025   &0.003    &0.060  &0.021  &0.003   &0.076     &0.030   &0.004    &0.050     &0.012   &0.002  \\

500,500  &0.050   &0.019   &0.002     &0.049  &0.018  &0.001   &0.060     &0.023   &0.003      &0.043    &0.013   &0.001  \\\hline

$\rho=0.5$  &   &   &   &  &   &    &     &   &  &    &  & \\



100,100   &0.029   &0.007   &0.001    &0.029  &0.013  &0.002   &0.061     &0.022   &0.003   &0.120     &0.077   &0.037  \\

100,200  &0.074   &0.028   &0.003    &0.069  &0.031  &0.007   &0.079   &0.031   &0.006   &0.055    &0.024   &0.010  \\

100,500  &0.119  &0.057   &0.014   &0.089  &0.042  &0.011   &0.093     &0.049   &0.011   &0.074     &0.032   &0.009\\

200,200   &0.084   &0.034   &0.009   &0.077  &0.036  &0.006   &0.094   &0.047   &0.011   &0.056    &0.024   &0.008  \\

200,500 &0.119   &0.061   &0.010   &0.086  &0.044  &0.010   &0.100     &0.037   &0.007   &0.080     &0.035   &0.008 \\

500,500  &0.130   &0.065   &0.011    &0.096  &0.044  &0.009   &0.102     &0.046   &0.009    &0.075     &0.031   &0.009  \\\hline

\end{tabular}}
\end{center}
\end{table}

\begin{table}[H]
\caption{Size of structural break tests with unknown break date in a factor model with Gaussian factors. $\epsilon=0.15$.}
\centering
\label{size_normal_tau015}
\begin{center}
\resizebox{\textwidth}{!}{
\begin{tabular}{l l r r r r r r r r r r r r r r r r r r r r r  r r r} \hline

$N,T$        &  \multicolumn{3}{c}{sup-Wald(HAC)}   &  \multicolumn{3}{c}{sup-LM(HAC)} &  \multicolumn{3}{c}{sup-LR} &\multicolumn{3}{c}{sup-LR$_m$}\\

              & 10\% & 5\%  &  1\%  & 10\% &  5\% &  1\%  &  10\%    & 5\%  & 1\% &  10\%    & 5\%  & 1\%  \\\hline

$\rho=0$&     &   &   &  &  &   &     &   &  &    &  &  \\

100,100     &0.009   &0.001   &0.000     &0.019  &0.007  &0.004   &0.056     &0.019   &0.002    &0.128     &0.086   &0.058  \\

100,200     &0.037   &0.013   &0.002     &0.038  &0.013  &0.002   &0.061     &0.027   &0.002   &0.040    &0.016   &0.009  \\

100,500    &0.062   &0.034   &0.004     &0.048  &0.024  &0.002   &0.064     &0.033   &0.006   &0.038     &0.015   &0.001 \\

200,200    &0.029   &0.012   &0.002    &0.026  &0.011  &0.002   &0.056     &0.019   &0.003  &0.048     &0.024   &0.010  \\

200,500   &0.061   &0.030   &0.005   &0.055  &0.022  &0.003   &0.073     &0.027   &0.003   &0.033     &0.009   &0.001  \\

500,500    &0.071   &0.030   &0.004   &0.056  &0.024  &0.004   &0.076     &0.032   &0.007    &0.040    &0.015   &0.002  \\\hline

$\rho=0.5$   &   &   &   &  &   &    &     &   &  &    &  &   \\

100,100   &0.017   &0.003   &0.001     &0.036  &0.021  &0.005   &0.068     &0.022   &0.005   &0.109     &0.073   &0.044 \\

100,200   &0.079   &0.032   &0.007    &0.072  &0.039  &0.010   &0.088    &0.04   &0.005   &0.065     &0.034   &0.012 \\

100,500    &0.154   &0.081   &0.017     &0.106  &0.053  &0.017   &0.108     &0.054   &0.013  &0.063     &0.031   &0.010 \\

200,200  &0.080   &0.032   &0.005     &0.072  &0.036  &0.013   &0.084     &0.030   &0.007   &0.052     &0.023   &0.006  \\

200,500   &0.156   &0.080   &0.022    &0.108  &0.056  &0.018   &0.102     &0.043   &0.011    &0.066     &0.027   &0.007  \\

500,500   &0.158   &0.084   &0.020   &0.121  &0.069  &0.020   &0.111     &0.059   &0.014   &0.066     &0.039   &0.014 \\\hline

\end{tabular}}
\end{center}
\end{table}

\begin{table}[H]
\caption{Size of structural break tests with unknown break date in a factor model with $t$-distributed factors. $\epsilon=0.15$.}
\centering
\label{size_t_10_tau015}
\begin{center}
\resizebox{\textwidth}{!}{
\begin{tabular}{l l r r r r r r r r r r r r r r r r r r r r r  r r r} \hline

$N,T$        &  \multicolumn{3}{c}{sup-Wald(HAC)}   &  \multicolumn{3}{c}{sup-LM(HAC)} &  \multicolumn{3}{c}{sup-LR} &    \multicolumn{3}{c}{sup-LR$_m$}\\

              & 10\% & 5\%  &  1\%  & 10\% &  5\% &  1\%  &  10\%    & 5\%  & 1\% &  10\%    & 5\%  & 1\%  \\\hline

$\rho=0$&       &   &   &  &  &   &     &   &  &    &  &  \\

100,100   &0.004   &0.002   &0.000     &0.015  &0.008  &0.003   &0.028     &0.007   &0.000    &0.128     &0.090   &0.061  \\

100,200   &0.038   &0.018   &0.002    &0.023  &0.008  &0.001   &0.048    &0.016   &0.002    &0.032     &0.012   &0.006  \\

100,500   &0.065   &0.028   &0.006    &0.051  &0.024  &0.002   &0.063     &0.023   &0.004    &0.028     &0.008   &0.000  \\

200,200  &0.031   &0.011   &0.001     &0.036  &0.011  &0.001   &0.045     &0.016   &0.002    &0.046     &0.022   &0.009 \\

200,500   &0.066   &0.025   &0.002    &0.056  &0.023  &0.003   &0.068     &0.028   &0.001   &0.031     &0.008   &0.001 \\

500,500  &0.069   &0.033   &0.005     &0.056  &0.027  &0.006   &0.064     &0.029   &0.002   &0.030    &0.008   &0.001  \\\hline

$\rho=0.5$  &   &   &   &  &   &     &     &   &  &    &  &  \\

100,100   &0.021   &0.008   &0.000    &0.037  &0.020  &0.007   &0.053     &0.021  &0.003    &0.107     &0.073   &0.048 \\

100,200   &0.068   &0.029   &0.006     &0.084  &0.046  &0.014   &0.085     &0.038   &0.008    &0.052     &0.030   &0.013  \\

100,500   &0.151   &0.087   &0.023    &0.101  &0.047  &0.015   &0.093     &0.043   &0.009    &0.054     &0.021   &0.006  \\

200,200   &0.071   &0.030   &0.006     &0.082  &0.043  &0.010   &0.066     &0.028   &0.003    &0.050     &0.024   &0.008  \\

200,500  &0.150   &0.085   &0.019    &0.115  &0.063  &0.023   &0.090     &0.041   &0.007    &0.051     &0.022   &0.005\\

500,500   &0.156   &0.085   &0.024     &0.108  &0.066  &0.021   &0.090     &0.045   &0.008    &0.059     &0.027   &0.006 \\\hline

\end{tabular}}
\end{center}
\end{table}

\begin{table}[H]
\caption{Power of structural break tests with unknown break date. $\epsilon=0.30$.}
\centering
\label{power_mean_shift_Normal_increase_samples}
\begin{center}
\resizebox{\textwidth}{!}{
\begin{tabular}{l l r r r r r r r r r r r r r r r r r r r r r  r r r} \hline

$N,T$        &  \multicolumn{3}{c}{sup-Wald(HAC)}  &  \multicolumn{3}{c}{sup-LM(HAC)} &  \multicolumn{3}{c}{sup-LR}&  \multicolumn{3}{c}{sup-LR$_m$}\\

             & 10\% & 5\%  &  1\%  & 10\% &  5\% &  1\%  &  10\%    & 5\%  & 1\% &  10\%    & 5\%  & 1\%  \\\hline

$\lambda_{i2}=\lambda_{i1}+\varepsilon_i$, $\varepsilon_i \overset{\mathrm{i.i.d.}}{\sim}N(0,I_{r_0})$ &  &   &   &  &  &   &     &   &  &    &  &  \\ \hline

$\rho=0$ &     &   &   &  &  &   &     &   &  &    &  &  \\



100,100    &0.651   &0.498   &0.259     &0.325  &0.254  &0.135   &1.000     &1.000   &0.998  &1.000     &0.999   &0.937  \\

100,200    &0.995   &0.989   &0.878     &0.880  &0.729  &0.503   &1.000     &1.000   &1.000   &1.000     &1.000   &0.994 \\

100,500     &1.000    &1.000    &1.000     &1.000   &1.000   &0.998   &1.000    &1.000    &1.000      &1.000    &1.000   &1.000   \\

200,200  &0.998   &0.992   &0.880    &0.876  &0.720  &0.505   &1.000     &1.000   &1.000    &1.000     &1.000   &1.000 \\

200,500    &1.000   &1.000   &1.000     &1.000  &1.000  &0.999   &1.000     &1.000   &1.000    &1.000   &1.000   &1.000   \\

500,500   &1.000   &1.000   &1.000    &1.000  &0.999  &0.998   &1.000     &1.000   &1.000    &1.000   &1.000   &1.000    \\\hline

$\rho=0.5$   &   &   &   &  &   &   &     &   &  &    &  &    \\



100,100    &0.614   &0.467   &0.229    &0.322  &0.256  &0.139   &1.000     &1.000   &0.999   &1.000     &0.998   &0.869  \\

100,200    &0.998   &0.984   &0.792   &0.861  &0.704  &0.424   &1.000    &1.000   &1.000    &1.000     &1.000   &0.997  \\

100,500    &1.000    &1.000    &1.000     &1.000   &1.000   &1.000   &1.000    &1.000    &1.000     &1.000    &1.000   &1.000   \\

200,200    &1.000    &0.980   &0.784     &0.860  &0.702  &0.419   &1.000     &1.000   &1.000    &1.000     &1.000   &0.999  \\

200,500   &1.000   &1.000   &1.000   &1.000  &1.000  &0.998   &1.000     &1.000   &1.000    &1.000     &1.000   &1.000  \\

500,500  &1.000   &1.000   &1.000    &1.000  &1.000  &1.000   &1.000     &1.000   &1.000    &1.000     &1.000   &1.000  \\\hline

\end{tabular}}
\end{center}
\end{table}

\begin{table}[H]
\caption{Power of structural break tests with unknown break date as the magnitude of the break in factor loadings increases. $\epsilon=0.15$.}
\centering
\label{power_mean_shift_Normal_increase_break_magnitude}
\begin{center}
\resizebox{\textwidth}{!}{
\begin{tabular}{l l r r r r r r r r r r r r r r r r r r r r r  r r r} \hline

$N,T$       &  \multicolumn{3}{c}{sup-Wald(HAC)}   &  \multicolumn{3}{c}{sup-LM(HAC)} &  \multicolumn{3}{c}{sup-LR}&   \multicolumn{3}{c}{sup-LR$_m$}\\

             & 10\% & 5\%  &  1\%  & 10\% &  5\% &  1\%  &  10\%    & 5\%  & 1\% &  10\%    & 5\%  & 1\%  \\\hline

$\lambda_{i2}=\lambda_{i1}+\varepsilon_i$, $\varepsilon_i \overset{\mathrm{i.i.d.}}{\sim}N(0,b\cdot I_{r_0})$   &  &   &   &  &  &   &     &   &  &     &   & \\
$N=100,T=100,\rho=0$   &  &   &   &  &  &   &     &   &  &     &   & \\
$b$ &      &   &   &  &  &   &     &   &  &    &  & \\

$0.0$  & 0.015     &0.005   &0.000   &0.026  &0.007  &0.002   &0.055     &0.018   &0.003  &0.158    &0.111  &0.074 \\

$0.2$ &0.036      &0.011   &0.001   &0.058  &0.022  &0.003   &0.098     &0.034   &0.004  &0.139    &0.090  &0.059 \\

$0.4$ &0.221      &0.128   &0.043   &0.184  &0.108  &0.039   &0.413     &0.320   &0.232  &0.330    &0.270  &0.140 \\

$0.6$ &0.561      &0.388   &0.171   &0.321  &0.223  &0.124   &0.917     &0.883   &0.844  &0.865    &0.834  &0.507 \\

$0.8$ &0.568      &0.424   &0.203   &0.290  &0.232  &0.145   &0.996     &0.994   &0.983  &0.990    &0.981  &0.715 \\

$1$   &0.530   &0.385   &0.188    &0.260  &0.216  &0.133   &1.000      &1.000    &0.996  &1.000     &0.997   &0.841 \\ \hline

$\lambda_{i2}=\lambda_{i1}+\varepsilon_i$, $\varepsilon_i \overset{\mathrm{i.i.d.}}{\sim}N(0,b\cdot I_{r_0})$   &  &   &   &  &  &   &     &   &  &     &   & \\
$N=100,T=200,\rho=0$   &  &   &   &  &  &   &     &   &  &     &   & \\
$b$    &   &   &   &  &   &   &      &  &  &     &   & \\

$0.0$  &0.039   &0.017   &0.003     &0.047  &0.020  &0.003   &0.075     &0.028   &0.003    &0.057     &0.026   &0.012  \\

$0.2$ &0.130      &0.061   &0.004   &0.162  &0.088  &0.016   &0.192     &0.099   &0.021  &0.089    &0.041  &0.010 \\

$0.4$ &0.755      &0.689   &0.560   &0.749  &0.634  &0.378   &0.809     &0.744   &0.642  &0.686    &0.625  &0.561 \\

$0.6$ &0.993      &0.971   &0.803   &0.77  &0.660  &0.432   &0.999     &0.998   &0.994  &0.997    &0.992  &0.972 \\

$0.8$ &0.996      &0.973   &0.800   &0.762  &0.629  &0.407   &1.000     &1.000   &1.000  &1.000    &1.000  &0.982 \\

$1$   &0.993   &0.967   &0.797   &0.749  &0.619  &0.411   &1.000     &1.000   &1.000  &1.000     &1.000   &0.990  \\\hline

$\lambda_{i2}=\lambda_{i1}+\varepsilon_i$, $\varepsilon_i \overset{\mathrm{i.i.d.}}{\sim}N(0,b\cdot I_{r_0})$   &  &   &   &  &  &   &     &   &  &     &   & \\
$N=100,T=100,\rho=0.5$   &  &   &   &  &  &   &     &   &  &     &   & \\
$b$   &   &   &   &  &   &   &  &  &  &    &  &  \\

$0.0$  &0.029      &0.011   &0.000   &0.070  &0.027  &0.006   &0.074     &0.031   &0.007  &0.150    &0.101  &0.056 \\

$0.2$  &0.054      &0.017   &0.002   &0.106  &0.047  &0.007   &0.115     &0.051   &0.009  &0.138    &0.086  &0.052 \\

$0.4$  &0.204      &0.117   &0.035   &0.184  &0.096  &0.035   &0.393     &0.318   &0.255  &0.338    &0.271  &0.133 \\

$0.6$  &0.484      &0.336   &0.130   &0.281  &0.186  &0.095   &0.885     &0.860   &0.827  & 0.835    &0.771  &0.420 \\

$0.8$  &0.529      &0.374   &0.170   &0.284  &0.211  &0.132   &0.990     &0.986   &0.978  &0.982    &0.948  &0.622 \\

$1.0$   &0.522   &0.372   &0.183    &0.256  &0.195  &0.130   &1.000     &1.000   &0.998    &1.000     &0.988   &0.747  \\ \hline

$\lambda_{i2}=\lambda_{i1}+\varepsilon_i$, $\varepsilon_i \overset{\mathrm{i.i.d.}}{\sim}N(0,b\cdot I_{r_0})$   &  &   &   &  &  &   &     &   &  &     &   & \\
$N=100,T=200,\rho=0.5$   &  &   &   &  &  &   &     &   &  &     &   & \\
$b$ &    &   &   &  &   &   &  &  &    &     &   &  \\

$0.0$  &0.076   &0.030   &0.003     &0.086  &0.044  &0.007   &0.080     &0.037   &0.002  &0.072     &0.034   &0.008  \\

$0.2$  &0.159      &0.071   &0.015   &0.192  &0.095  &0.025   &0.185     &0.094   &0.020  &0.085    &0.040  &0.015 \\

$0.4$  &0.718      &0.644   &0.469   &0.681  &0.550  &0.276   &0.757     &0.698   &0.612  &0.641    &0.594  &0.536 \\

$0.6$  &0.986      &0.922   &0.685   &0.756  &0.624  &0.382   &0.997     &0.996   &0.991  &0.990    &0.983  &0.953 \\

$0.8$  &0.986      &0.931   &0.682   &0.715  &0.556  &0.35   &1.000     &1.000   &1.000  &1.000    &1.000  &0.979 \\

$1$   &0.994   &0.932   &0.721   &0.723  &0.568   &0.366     &1.000   &1.000  &1.000       &1.000   &1.000 &0.980    \\\hline

\end{tabular}}
\end{center}
\end{table}

\begin{table}[H]
\caption{Power of structural break tests with unknown break date under rotational changes. $\epsilon=0.15$.}
\centering
\label{power_rotational_change_increase_aa}
\begin{center}
\resizebox{\textwidth}{!}{
\begin{tabular}{l l r r r r r r r r r r r r r r r r r r r r r  r r r} \hline

$N,T$       &  \multicolumn{3}{c}{sup-Wald(HAC)} &  \multicolumn{3}{c}{sup-LM(HAC)} &  \multicolumn{3}{c}{sup-LR} &\multicolumn{3}{c}{sup-LR$_m$}\\

            & 10\% & 5\%  &  1\%  & 10\% &  5\% &  1\%  &  10\%    & 5\%  & 1\% &  10\%    & 5\%  & 1\% \\\hline

$\Lambda_2=a\Lambda_1,N=T=100,\rho=0$&      &   &   &     &   &  &    &  &  &      &   & \\
$a$ &    &   &      &  &   &   &  &    &  &      &   & \\

1.0      &0.012   &0.002   &0.000     &0.012  &0.006  &0.003   &0.050     &0.018   &0.002    &0.145     &0.107  &0.067  \\

1.1    &0.009   &0.001   &0.000    &0.023  &0.007  &0.002   &0.054     &0.018   &0.003   &0.127     &0.086   &0.055 \\

1.2   &0.011   &0.004   &0.000    &0.043  &0.021  &0.003   &0.114     &0.038   &0.004   &0.096     &0.059   &0.030 \\

1.3   &0.029   &0.006   &0.000   &0.053  &0.022  &0.002   &0.252     &0.112   &0.009  &0.101     &0.051   &0.024 \\

1.4   &0.035   &0.010   &0.000    &0.079  &0.029  &0.004   &0.457     &0.241   &0.040    &0.121     &0.057   &0.020  \\

1.5   &0.060   &0.013   &0.002   &0.070  &0.028  &0.003   &0.653     &0.405   &0.100    &0.162     &0.073   &0.016 \\

1.6  &0.055   &0.013   &0.001     &0.072  &0.029  &0.006   &0.771     &0.538   &0.161  &0.192     &0.084   &0.021 \\

1.7    &0.076   &0.016   &0.001     &0.064  &0.025  &0.006   &0.904     &0.708   &0.253    &0.262     &0.119   &0.033 \\

1.8   &0.084   &0.013   &0.000    &0.048  &0.012  &0.004   &0.946     &0.800   &0.328    &0.301     &0.130   &0.034  \\

1.9  &0.097  &0.021   &0.001    &0.045  &0.012  &0.004   &0.976     &0.884   &0.378  &0.350     &0.145   &0.043  \\

2.0   &0.108   &0.016   &0.000   &0.047  &0.012  &0.001   &0.989     &0.919   &0.460   &0.404     &0.161   &0.049 \\\hline

$\Lambda_2=a\Lambda_1,N=T=100,\rho=0.5$&        &   &  &    &  &  &     &   & &     &   & \\
$a$ &    &   &    &   &   &   &      &     & &     &   & \\

1.0    &0.026   &0.008   &0.002    &0.044  &0.021  &0.006   &0.065     &0.030   &0.007    &0.120     &0.086   &0.043  \\

1.1  &0.024   &0.009   &0.001    &0.044  &0.018  &0.004   &0.071     &0.028   &0.004    &0.100     &0.068   &0.040  \\

1.2    &0.025   &0.009   &0.001   &0.055  &0.028  &0.010   &0.124     &0.045   &0.008    &0.104     &0.071   &0.040  \\

1.3    &0.029   &0.011   &0.001    &0.058  &0.025  &0.011   &0.178     &0.082   &0.015    &0.094     &0.060   &0.031  \\

1.4  &0.049   &0.016   &0.003    &0.071  &0.034  &0.007   &0.292     &0.144   &0.028   &0.073     &0.041   &0.020 \\

1.5  &0.042   &0.015   &0.003    &0.073  &0.034  &0.008   &0.433     &0.244   &0.056   &0.114     &0.057   &0.023\\

1.6   &0.048   &0.020   &0.001     &0.067  &0.026  &0.008   &0.567     &0.352   &0.103    &0.122     &0.053   &0.023  \\

1.7  &0.063   &0.019   &0.003    &0.053  &0.023  &0.007   &0.687     &0.456   &0.142   &0.166     &0.080   &0.029 \\

1.8   & 0.061   &0.014   &0.002     &0.048  &0.016  &0.002   &0.794     &0.557   &0.162    &0.192     &0.087   &0.033 \\

1.9  &0.065   &0.017   &0.002    &0.044  &0.012  &0.003   &0.865     &0.656   &0.218    &0.251     &0.117   &0.034  \\

2.0  &0.079   &0.020   &0.000    &0.039  &0.009  &0.001   &0.916     &0.743   &0.270   &0.294     &0.127   &0.030  \\\hline

$\Lambda_2=a\Lambda_1,N=100,T=200,\rho=0$&        &   &  &    &  &  &     &   & &     &   & \\
$a$&    &   &    &   &   &   &        &   & &     &   & \\
1.0  &0.032   &0.014   &0.001     &0.024  &0.007 &0.001   &0.061     &0.026   &0.001  &0.048     &0.025   &0.011  \\

1.1  &0.060   &0.026   &0.005     &0.060  &0.026  &0.003   &0.117     &0.046  &0.002  &0.053     &0.021   &0.008  \\

1.2  &0.179   &0.083   &0.012     &0.207  &0.101  &0.017   &0.368     &0.213   &0.048  &0.120     &0.043   &0.007 \\

1.3   &0.444   &0.225   &0.027     &0.428  &0.219  &0.039   &0.756     &0.591   &0.241   &0.309     &0.129   &0.017 \\

1.4  &0.704   &0.446   &0.074    &0.612  &0.347  &0.068   &0.948     &0.871   &0.568  &0.532     &0.269   &0.067  \\

1.5  &0.867   &0.635   &0.111     &0.714  &0.383 &0.066   &0.996     &0.977   &0.770  &0.700     &0.353   &0.094  \\

1.6  &0.951   &0.779   &0.175    &0.768  &0.408  &0.053   & 1.000    &0.998   &0.911  &0.816     &0.453   &0.112  \\

1.7  &0.983   &0.858   &0.193    &0.807  &0.404  &0.033   &1.000     &1.000   &0.962 &0.879     &0.505   &0.110 \\

1.8  &0.991   &0.901   &0.221     &0.833  &0.400  &0.025   &1.000     &1.000   &0.984  &0.923     &0.563   &0.100 \\

1.9 &0.994   &0.934   &0.252      &0.852  &0.409  &0.012   &1.000     &1.000   &0.993  &0.961     &0.611   &0.078  \\

2.0  &0.997   &0.952   &0.260   &0.867  &0.376  &0.014   &1.000     &1.000   &0.997 &0.975     &0.685   &0.079  \\\hline

$\Lambda_2=a\Lambda_1,N=100,T=200,\rho=0.5$&         &   &  &    &  &  &     &   & &     &   & \\
$a$ &    &   &    &   &   &   &  &   &   &         &   & \\

1.0   &0.080   &0.034   &0.005     &0.081  &0.041  &0.012   &0.075     &0.030   &0.007    &0.067     &0.034   &0.011  \\

1.1   &0.090   &0.035   &0.007     &0.094  &0.048  &0.012   &0.123     &0.052   &0.011  &0.060     &0.029   &0.010  \\

1.2   &0.152   &0.065   &0.013    &0.179  &0.090  &0.022   &0.250     &0.125   &0.033    &0.085     &0.041   &0.016  \\

1.3    &0.280  &0.137   &0.023   &0.290  &0.149  &0.036   &0.507     &0.333   &0.096  &0.141     &0.069   &0.023\\

1.4   &0.446   &0.230   &0.037    &0.380  &0.175  &0.033   &0.759     &0.579   &0.225    &0.228     &0.103   &0.030  \\

1.5  &0.609  &0.327   &0.049   &0.481  &0.212  &0.023   &0.909     &0.789   &0.420  &0.385     &0.182   &0.045  \\

1.6   &0.726   &0.439   &0.065    &0.557  &0.246  &0.015   &0.974     &0.908   &0.578    &0.502     &0.256   &0.072  \\

1.7  &0.846   &0.574   &0.079    &0.646  &0.273  &0.016   &0.996     &0.971   &0.756    &0.652     &0.355   &0.106  \\

1.8  &0.887   &0.620   &0.081   &0.662  &0.264  &0.017   &0.999     &0.985   &0.829   &0.751     &0.405   &0.099  \\

1.9  &0.931   &0.710   &0.097    &0.705  &0.271  &0.016   &0.999     &0.996   &0.912    &0.839     &0.493   &0.106  \\

2.0   &0.949   &0.736   &0.089    &0.741  &0.278  &0.003   &1.000     &0.998   &0.937    &0.885     &0.579   &0.127  \\\hline

\end{tabular}}
\end{center}
\end{table}

\begin{table}[H]
\caption{Power of structural break tests with unknown break date in the case of disappearing factors. $\epsilon=0.15$.}
\centering
\label{power_singular_rotational_change_increase_samples}
\begin{center}
\resizebox{\textwidth}{!}{
\begin{tabular}{l l r r r r r r r r r r r r r r r r r r r r r  r r r} \hline

$N,T$       &  \multicolumn{3}{c}{sup-Wald(HAC)}  &  \multicolumn{3}{c}{sup-LM(HAC)} &  \multicolumn{3}{c}{sup-LR} &\multicolumn{3}{c}{sup-LR$_m$}\\

             & 10\% & 5\%  &  1\%  & 10\% &  5\% &  1\%  &  10\%    & 5\%  & 1\% &  10\%    & 5\%  & 1\% \\\hline

$u_t\sim N(0,1-\rho^2)$     &  &   &   &     &  &  &     &   & &     &   & \\

100,100   &0.265  &0.144   &0.033      &0.199  &0.114  &0.041   &0.999     &0.995   &0.934   &0.926     &0.755   &0.365  \\

100,200   &0.896   &0.738   &0.397   &0.772  &0.530  &0.211   &1.000     &1.000   &1.000    &0.998     &0.988   &0.898  \\

100,500   &1.000   &1.000   &0.999    &1.000  &1.000  &0.975   &1.000     &1.000   &1.000   &1.000     &1.000   &0.998  \\

200,200   &0.904   &0.775   &0.401     &0.792  &0.542  &0.207   &1.000     &1.000   &1.000    &1.000     &0.999   &0.982  \\

200,500   &1.000   &1.000   &0.999    &1.000  &1.000  &0.978   &1.000     &1.000   &1.000    &1.000     &1.000   &1.000  \\

500,500  &1.000   &1.000   &0.998     &1.000  &1.000  &0.978   &1.000     &1.000   &1.000    &1.000     &1.000   &1.000  \\\hline

$u_t\sim \sqrt{\frac{1-\rho^2}{1.25}}t(10)$    &     &   &  &    &  &  &     &   & &     &   & \\

100,100    &0.243   &0.127   &0.029    &0.210  &0.123  &0.047   &0.995     &0.986   &0.897    &0.893     &0.722   &0.352  \\

100,200   &0.824   &0.668   &0.324   &0.708  &0.474  &0.191   &1.000     &1.000   &0.999   &0.998     &0.983   &0.869  \\

100,500   &1.000   &0.998   &0.985    &0.998  &0.996  &0.934   &1.000     &1.000   &1.000  &1.000     &1.000   &0.999  \\

200,200  &0.844   &0.677   &0.333   &0.736  &0.498  &0.187   &1.000     &1.000   &0.999    &1.000     &0.998   &0.956  \\

200,500   &0.999   &0.999   &0.984    &0.999  &0.997  &0.951   &1.000     &1.000   &1.000    &1.000     &1.000   &1.000  \\

500,500   &1.000   &0.999   &0.985   &0.998  &0.995  &0.934   &1.000     &1.000   &1.000   &1.000     &1.000   &1.000  \\\hline

\end{tabular}}
\end{center}
\end{table}

\begin{table}[H]
\caption{Power of structural break tests with unknown break date in the case of factor mean change. $\epsilon=0.15$.}
\centering
\label{power_mean_shift_change_increase_samples_dd}
\begin{center}
\resizebox{\textwidth}{!}{
\begin{tabular}{l l r r r r r r r r r r r r r r r r r r r r r  r r r} \hline

$N,T$       &  \multicolumn{3}{c}{sup-Wald(HAC)}  &  \multicolumn{3}{c}{sup-LM(HAC)} &  \multicolumn{3}{c}{sup-LR}&   \multicolumn{3}{c}{sup-LR$_m$}\\

            & 10\% & 5\%  &  1\%  & 10\% &  5\% &  1\%  &  10\%    & 5\%  & 1\% &  10\%    & 5\%  & 1\%\\\hline

$\rho=0$  &   &  &   &     &    &  &     &   & &     &   & \\

100,100 &0.048   &0.014  &0.000   &0.071     &0.026    &0.005  &0.128     &0.049   &0.005 &0.196     &0.106   &0.037 \\

100,200 &0.202    &0.103  &0.018   &0.215     &0.109    &0.022  &0.269     &0.141   &0.028 &0.551     &0.335   &0.101 \\

100,500 &0.653    &0.499  &0.221   &0.649     &0.508    &0.222  &0.678     &0.531   &0.256 &0.967     &0.919   &0.611 \\

200,200 &0.199   &0.096  &0.020   &0.215     &0.107    &0.019  &0.277     &0.149   &0.032 &0.556     &0.341   &0.115 \\

200,500 &0.651    &0.501  &0.245   &0.644     &0.511    &0.239  &0.668     &0.525   &0.271 &0.974     &0.917   &0.590 \\

500,500 &0.638     &0.501  &0.241   &0.639     &0.508    &0.228  &0.669     &0.526  &0.267 &0.982     &0.932   &0.607 \\

1000,1000 &0.957    &0.915  &0.771   &0.954     &0.910    &0.748  &0.951     &0.917   &0.771 &1.000     &0.998   &0.985 \\\hline

$\rho=0.5$  &     &  &   &     &    &  &     &   & &     &   & \\

100,100    &0.071  &0.030   &0.003    &0.103     &0.054    &0.016  &0.118     &0.052   &0.010 &0.175     &0.097   &0.040 \\

100,200  &0.164  &0.086   &0.019   &0.187  &0.108     &0.029  &0.186     &0.095   &0.019 &0.322     &0.171   &0.050 \\

100,500  &0.449  &0.302   &0.110   &0.409  &0.260   &0.097  &0.417     &0.275   &0.091 &0.842     &0.677   &0.305 \\

200,200   &0.164  &0.080   &0.022    &0.191     &0.111    &0.024  &0.177     &0.093   &0.025 &0.304     &0.167   &0.042 \\

200,500 &0.468 &0.315   &0.115   &0.428  &0.300  &0.122   &0.435      &0.295   &0.103 &0.836     &0.679   &0.299 \\

500,500   &0.464  &0.308   &0.112     &0.407     &0.263    &0.090  &0.425     &0.285   &0.099 &0.833     &0.661   &0.314 \\

1000,1000  &0.757  &0.646   &0.375    &0.731     &0.596    &0.326  &0.754     &0.637   &0.349 &0.997     &0.990   &0.876 \\

\hline
\end{tabular}}
\end{center}
\end{table}

\section{Empirical application}

In this section, we estimate a factor model for the US industrial employment rates and apply the proposed tests to check whether the
factor loadings  have undergone a structural change in the past decade. Monthly data from January 2010 through April 2022 are available from the US Department of Labor for 84 industries. The data are a balanced panel with $T=147,N=84$.

We use the information criteria $IC_{p1}$ and $IC_{p2}$ of \cite{Bai2002}, the $ER$ (eigenvalue ratio) and $GR$ (growth ratio) of \cite{Ahn2013}, the empirical distribution estimator of \cite{Onatski2010}, and the bridge estimator of \cite{Caner2014} to determine the number of common factors in the data. The maximum number of factors is set to 10. The estimated number of factors by $IC_{p1}$ and $IC_{p2}$ is three; the methods of \cite{Onatski2010} and \cite{Caner2014} detect two factors; and $ER$ and $GR$ of \cite{Ahn2013} choose one common factor. Therefore, we study the test results for the cases of one, two, and three common factors.

We apply our LR and LR$_m$ tests to examine whether there exists a structural break in factor loadings and estimate the break date if the null hypothesis is rejected. All of the settings for the two tests are the same as those used in the simulation studies.

Table \ref{empirical_application_reject} reports the results of the LR and LR$_m$ tests at the $1\%$, $5\%$, and $10\%$ significance levels. Both the sup-LR and the sup-LR$_m$ reject the null hypothesis for $r=1,2,3$. We also use the sup-LR and sup-LR$_m$ to estimate the break date. The sup-LR and sup-LR$_m$ show that
the break date occurs in January 2020 ($\hat k=121$), regardless of the number of factors. We employed Wald and LM tests, both of which rejected the null hypothesis. Furthermore, we utilized a sequential procedure and determined that there was only one breakpoint present within the dataset.

\begin{figure}[H]
  \centering
  \includegraphics[width=1.0\textwidth]{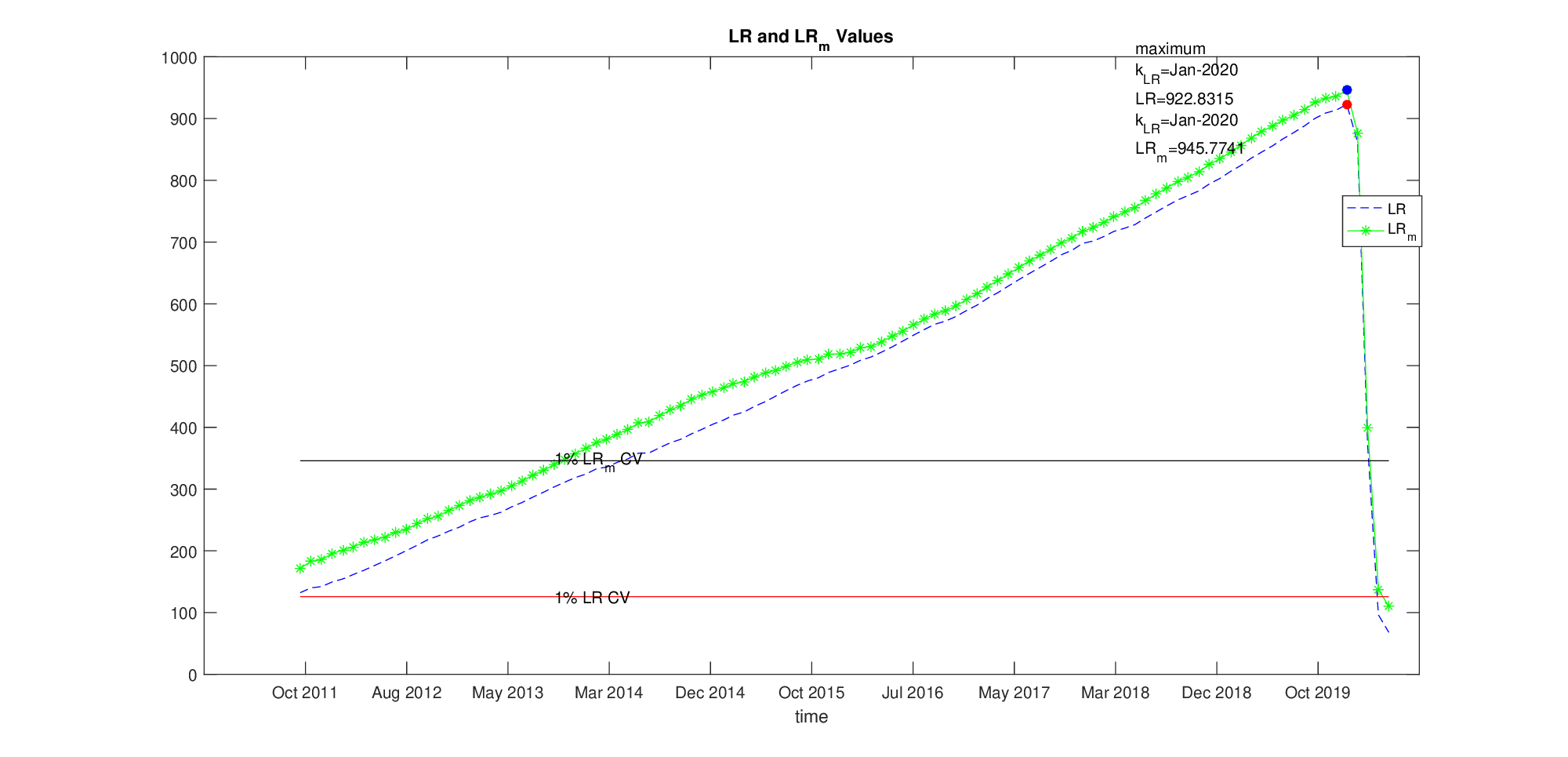}\\
  \caption{LR and LR$_m$ values with different $k$ when $r=3$. The vertical axis denotes the LR and LR$_m$ values, and the horizontal axis corresponds to the different break date $k$. The red and black horizontal lines are the $1\%$ sup-LR critical value and the $1\%$ sup-LR$_m$ critical value, respectively. The green and blue curves are the values of
LR and LR$_m$, respectively. The red and blue points are the maximum points of LR and LR$_m$, respectively. }\label{empirical_QML_value}
\end{figure}

Figure \ref{empirical_QML_value} shows the values of LR and LR$_m$ with different $k$. The red and black horizontal lines are the
$1\%$ sup-LR critical value and the $1\%$ sup-LR$_m$ critical value, respectively.
The green and blue curves are the values of the LR and the LR$_m$ statistics, respectively.
The red and blue points are the maximum points of the LR and the LR$_m$, respectively. The break date in January 2020 means that employment has been changed to a new
regime since February 2020, which is consistent
with the observation that the stock market also fell
sharply in late February.

We also separately performed LR and LR$_m$ tests on each single factor and found
that all of them rejected the null hypothesis at the 1$\%$, 5$\%$, and 10$\%$ significance levels, except for the LR$_m$ test on the third factor at the 1$\%$ and 5$\%$ significance levels. The breakpoint positions for these single factors were identified at January 2020, February 2020, and March 2020, respectively.
To explore the variations in factor contribution over time, we calculated the proportion of explained variation by summing the first three eigenvalues of $X_{\hat k}^{(i)}  X_{\hat k}^{(i)'}$ and dividing it by the trace of $X_{\hat k}^{(i)}  X_{\hat k}^{(i)'}$, $i=1,2$ for data before and after the breakpoint. The results showed a significant increase in the proportion of explained variation value, from $0.227$ before the breakpoint to $0.428$ after the breakpoint, indicating a stronger degree of co-movement of employment across different industries during the pandemic.

\begin{table}[H]
\caption{Tests for the structural change in industrial employment rates in the US from January 2010 to April 2022.}
\centering
\label{empirical_application_reject}
\begin{center}
\begin{tabular}{l r r r r r r r r r r r r r r r r r r r r r r  r r r} \hline

      & \multicolumn{3}{c}{sup-LR}&   \multicolumn{3}{c}{sup-LR$_m$}\\

            &  10\%  &  5\% &  1\%  &   10\%  &  5\% &  1\% \\\hline

$r=1$ &1    &1   &1    &1   &1   &1 \\

$r=2$ &1    &1   &1    &1  &1   &1 \\

$r=3$ &1    &1   &1    &1   &1   &1 \\
\hline

\end{tabular}

        \footnotesize
     \item[*] 1 means rejection.
\end{center}
\end{table}

\section{Conclusions}

This paper proposes an LR test for structural changes in the factor loading matrix. The new test is based on the LR principle under the assumption that the factors are normally
distributed and serially uncorrelated. We show that the limiting null distribution of the test statistic is a function of a Brownian bridge that depends on a long-run variance term, which allows for non-normal
and serially dependent factors. Under the alternative hypothesis, the test statistic diverges at a faster rate than regular Wald-type tests if the pseudo-factors have a singular variance before or after the
break. The new LR tests are also generalized to allow for factor mean changes and multiple breaks. The simulation results confirm that our test is much more powerful than the Wald test of HI
(2015). The test procedure is illustrated with a monthly US employment dataset and detects a structural break in early 2020.

\section*{\textit{Appendix}}

\textit{\large \setcounter{equation}{0} \setcounter{subsection}{0}
\setcounter{equation}{0} \setcounter{proposition}{0}}{\large \par}

\subsection*{The LR test is invariant to the identification conditions}

The identification problem in factor models occurs because one can set
$G\Lambda^{\prime}=\bar{G}\bar{\Lambda}^{\prime}$, where $\bar{G}=GQ$
and $\bar{\Lambda}^{\prime}=Q^{-1}\Lambda^{\prime}$ for any nonsingular
matrix $Q$. The identification condition essentially determines the
rotation matrix $Q$.

To understand why the LR test is invariant to the choice of identification
conditions, let $\hat{g}_{t}$ denote the estimated factors under
condition (3.1) and $\tilde{g}_{t,Q}$ denote the factors estimated
under an alternative identification condition. There must exist a
nonsingular matrix $Q$ such that \[
\tilde{g}_{t,Q}=Q^{\prime}\hat{g}_{t}.\]
Thus, under the alternative identidication condition, the pre-$k$
and post-$k$ factor variances are given by \begin{eqnarray*}
\tilde{\Sigma}_{1}(k) & = & \frac{1}{k}\sum_{t=1}^{k}\tilde{g}_{t,Q}\tilde{g}_{t,Q}^{\prime}=Q^{\prime}\hat{\Sigma}_{1}(k)Q\\
\tilde{\Sigma}_{2}(k) & = & \frac{1}{T-k}\sum_{t=k+1}^{T}\tilde{g}_{t,Q}\tilde{g}_{t,Q}^{\prime}=Q^{\prime}\hat{\Sigma}_{2}(k)Q.\end{eqnarray*}
The quasi-Gaussian likelihood in (3.2) for a break point at $k$ is
given by\begin{eqnarray*}
\mathcal{L}_{Q}(k) & = & k\log(|\tilde{\Sigma}_{1}(k)|)+(T-k)\log(|\tilde{\Sigma}_{2}(k)|)\\
 & = & k\log(|Q^{\prime}|\cdot|\hat{\Sigma}_{1}(k)|\cdot|Q|)+(T-k)\log(|Q^{\prime}|\cdot|\hat{\Sigma}_{2}(k)|\cdot|Q|)\\
 & = & 2T\log(|Q|)+k\log(|\hat{\Sigma}_{1}(k)|)+(T-k)\log(|\hat{\Sigma}_{2}(k)|)\end{eqnarray*}
The log-likelihood of no change for the entire sample becomes\[
\mathcal{L}_{Q0}=T\log(|T^{-1}\sum_{t=1}^{T}\tilde{g}_{t,Q}\tilde{g}_{t,Q}^{\prime}|)=T\log(|Q^{\prime}|\cdot|T^{-1}\sum_{t=1}^{T}\hat{g}_{t}\hat{g}_{t}^{\prime}|\cdot|Q|)=2T\log(|Q|)\]
because $T^{-1}\sum_{t=1}^{T}\hat{g}_{t}\hat{g}_{t}^{\prime}=I_{r}$
under (3.1). Thus, the likelihood ratio test under this alternative
identification condition can be expressed as \begin{eqnarray*}
LR_{Q}(k) & \equiv & -k\log(|\tilde{\Sigma}_{1}(k)|)-(T-k)\log(|\tilde{\Sigma}_{2}(k)|)+\mathcal{L}_{Q0}\\
 & = & -k\log(|\hat{\Sigma}_{1}(k)|)-(T-k)\log(|\hat{\Sigma}_{2}(k)|)\\
 & \equiv & LR(k).\end{eqnarray*}
Thus, the above $LR_{Q}(k)$ is equivalent to the $LR(k)$.

\begin{lemma}\label{eq:Bai implied}
Under Assumptions \ref{factors}-\ref{invariance},

(i).\begin{eqnarray}
\max_{[\epsilon T]\le k\le[(1-\epsilon)T]}\|\frac{1}{k}\sum\limits _{t=1}^{k}(\hat{g}_{t}-H^{\prime}g_{t})\|=O_{p}(\delta_{NT}^{-2})\label{eq:need to prove}
\end{eqnarray}

(ii). \begin{eqnarray}
\max_{[\epsilon T]\le k\le[(1-\epsilon)T]}\|k^{-1}\sum\limits _{t=1}^{k}(\hat{g}_{t}\hat{g}_{t}^{\prime}-H^{\prime}g_{t}g_{t}^{\prime}H)\| & =O_{p}(\delta_{NT}^{-2})\label{eq:delta_NT_uniform}\end{eqnarray}
\end{lemma}

\textbf{{Proof.}} See the online appendix. $\Box$\textit{\large \vspace{1em}
}{\large \par}

{\bf Proof of Theorem \ref{distribution_theorem}:}

Recall that $H=(\Lambda^{'}\Lambda/N)(G^{'}\hat{G}/T)V_{NT}^{-1}$
and its probability limit is \[
H_{0}=\mathrm{plim}_{N,T\to\infty}H.\]
By (\ref{eq:HggH=I}), we have \begin{equation}
H_{0}^{\prime}\Sigma_{G}H_{0}=I_{r}.\label{eq:HSH=I}\end{equation}
Now, note that

\begin{align}
I_{r}=T^{-1}\hat{G}^{\prime}\hat G=T^{-1}H^{\prime}G^{\prime}GH+O_{p}(\delta_{NT}^{-2})\label{eq:delta_NT}\end{align}
by Lemmas B2 and B3 of Bai (2003).
Let $R=\Sigma_{G}^{1/2}H_{0}$ (i.e., $R$ is an orthogonal matrix),
so we have \begin{align}
&T^{-1/2}\sum\limits _{t=1}^{k}(\hat{g}_{t}\hat{g}_{t}'-I_{r}) \nonumber \\
& =\frac{1}{\sqrt{T}}\sum\limits _{t=1}^{k}H^{\prime}g_{t}g_{t}^{\prime}H-\frac{k}{\sqrt{T}}I_{r}+\sqrt{T}O_{p}(\delta_{NT}^{-2})\nonumber \\
 & =H^{\prime}\left(\frac{1}{\sqrt{T}}\sum\limits _{t=1}^{k}g_{t}g_{t}^{\prime}-\frac{k}{T}\frac{1}{\sqrt{T}}\sum\limits _{t=1}^{T}g_{t}g_{t}^{\prime}\right)H+\sqrt{T}O_{p}(\delta_{NT}^{-2})\nonumber \\
 & =H^{\prime}\Sigma_{G}^{1/2}\left(\frac{1}{\sqrt{T}}\sum\limits _{t=1}^{k}(\eta_{t}\eta_{t}^{\prime}-I_{r})-\frac{k}{T}\frac{1}{\sqrt{T}}\sum\limits _{t=1}^{T}(\eta_{t}\eta_{t}^{\prime}-I_{r})\right)\Sigma_{G}^{1/2}H+\sqrt{T}O_{p}(\delta_{NT}^{-2})\nonumber \\
 & =R^{\prime}\left[\frac{1}{\sqrt{T}}\sum\limits _{t=1}^{k}(\eta_{t}\eta_{t}^{\prime}-I_{r})-\frac{k}{T}\frac{1}{\sqrt{T}}\sum\limits _{t=1}^{T}(\eta_{t}\eta_{t}^{\prime}-I_{r})\right]R+\sqrt{T}O_{p}(\delta_{NT}^{-2})+o_{p}(1)\label{eq:gtgt-Ir expansion}\\
 & \Rightarrow R^{\prime}[\zeta(\pi)-\pi\zeta(1)]R\label{eq:gtgt-Ir}\end{align}
under the condition that $\sqrt{T}/N\to0$, where the first line uses
(\ref{eq:delta_NT_uniform}), the second line uses (\ref{eq:delta_NT}),
the fourth line follows from the definition of $R$, and the last
line follows from Assumption \ref{limit_distribution2}.

Because $T^{-1}\sum\limits _{t=1}^{T}\hat{g}_{t}\hat{g}_{t}^{\prime}=I_{r}$,
it follows that \begin{align}
T^{-1/2}\sum\limits _{t=k+1}^{T}(\hat{g}_{t}\hat{g}_{t}^{\prime}-I_{r}) & =-T^{-1/2}\sum\limits _{t=1}^{k}(\hat{g}_{t}\hat{g}_{t}^{\prime}-I_{r})\nonumber \\
 & \Rightarrow R^{\prime}[\zeta(\pi)-\pi\zeta(1)]R\label{eq:gtgt2}\end{align}
Moreover, (\ref{eq:gtgt-Ir}) and (\ref{eq:gtgt2}) imply that \begin{align}
\hat{\Sigma}_{1}(k)-I_{r} & =\frac{\sqrt{T}}{k}\frac{1}{\sqrt{T}}\sum\limits _{t=1}^{k}(\hat{g}_{t}\hat{g}_{t}^{\prime}-I_{r})=O_{p}(T^{-1/2})\nonumber \\
\hat{\Sigma}_{2}(k)-I_{r} & =\frac{\sqrt{T}}{T-k}\frac{1}{\sqrt{T}}\sum\limits _{t=k+1}^{T}(\hat{g}_{t}\hat{g}_{t}^{\prime}-I_{r})=O_{p}(T^{-1/2})\label{eq:Sig1-Ir}\end{align}
and the $O_{p}(T^{-1/2})$ terms are uniform over $[\epsilon T]\le k\le T-[\epsilon T]$
for $\epsilon\in(0,1)$.

Consider the second order Taylor expansions of $\log(|\hat{\Sigma}_{1}(k)|)$
and $\log(|\hat{\Sigma}_{2}(k)|)$ at $I_{r}$, so \begin{align*}
\log(|\hat{\Sigma}_{1}(k)|) & =\tr[\hat{\Sigma}_{1}(k)-I_{r}]-\frac{1}{2}\tr((\hat{\Sigma}_{1}(k)-I_{r})(\hat{\Sigma}_{1}(k)-I_{r})+o_{p}(T^{-1})\\
\log(|\hat{\Sigma}_{2}(k)|) & =\tr[\hat{\Sigma}_{2}(k)-I_{r}]-\frac{1}{2}\tr((\hat{\Sigma}_{2}(k)-I_{r})(\hat{\Sigma}_{2}(k)-I_{r})+o_{p}(T^{-1}),\end{align*}
where the $o_{p}(T^{-1})$ terms are uniform over $[\epsilon T]\le k\le T-[\epsilon T]$
and follow from (\ref{eq:Sig1-Ir}). Thus,\begin{align}
k\log(|\hat{\Sigma}_{1}(k)|) & =\tr\left(\sum\limits _{t=1}^{k}(\hat{g}_{t}\hat{g}_{t}'-I_{r})\right)-\frac{k}{2}\tr\left(\left[\frac{1}{k}\sum\limits _{t=1}^{k}(\hat{g}_{t}\hat{g}_{t}'-I_{r})\right]^{2}\right)+o_{p}(1)\nonumber \\
(T-k)\log(|\hat{\Sigma}_{2}(k)|) & =\tr\left(\sum\limits _{t=k+1}^{T}(\hat{g}_{t}\hat{g}_{t}'-I_{r})\right)-\frac{T-k}{2}\tr\left(\left[\frac{1}{T-k}\sum\limits _{t=k+1}^{T}(\hat{g}_{t}\hat{g}_{t}'-I_{r})\right]^{2}\right)+o_{p}(1),\label{eq:each term in LR}\end{align}
where the $o_{p}(1)$ terms are uniform over $[\epsilon T]\le k\le T-[\epsilon T]$.
Because $\sum\limits _{t=1}^{k}(\hat{g}_{t}\hat{g}_{t}'-I_{r})+\sum\limits _{t=k+1}^{T}(\hat{g}_{t}\hat{g}_{t}'-I_{r})=0$,
we have \begin{eqnarray}
LR(k,\hat{G}) & = & \frac{k}{2}\tr\left(\left[\frac{1}{k}\sum\limits _{t=1}^{k}(\hat{g}_{t}\hat{g}_{t}'-I_{r})\right]^{2}\right)+\frac{T-k}{2}\tr\left(\left[\frac{1}{T-k}\sum\limits _{t=k+1}^{T}(\hat{g}_{t}\hat{g}_{t}'-I_{r})\right]^{2}\right)+o_{p}(1)\nonumber \\
 & \Rightarrow & \frac{1}{2\pi(1-\pi)}\tr\left([\zeta(\pi)-\pi\zeta(1)]^{2}\right)\label{eq:LR distribution}\end{eqnarray}
by (\ref{eq:gtgt-Ir}), (\ref{eq:gtgt2}), and the fact that $R$ is
orthonormal. Because $\tr(AA)=\vec(A)'\vec(A)$ for a symmetric matrix
$A$, the result in (\ref{eq:LR distribution}) can be rewritten as
\begin{align}
 & \frac{1}{2\pi(1-\pi)}\vec(\zeta(\pi)-\pi\zeta(1))'\vec(\zeta(\pi)-\pi\zeta(1))\nonumber \\
\overset{\mathrm{d}}{=} & \frac{1}{2\pi(1-\pi)}[\mathbb{W}(\pi)-\pi\mathbb{W}(1)]'\Omega[\mathbb{W}(\pi)-\pi\mathbb{W}(1)],\label{eq:LR distribution trace}\end{align}
where $\overset{\mathrm{d}}{=}$ means equality in distribution only,
\[
\Omega=\mathrm{plim}_{T\to\infty}\var[\frac{1}{\sqrt{T}}\sum_{t=1}^{T}\vec(\Sigma_{G}^{-1/2}g_{t}g_{t}^{\prime}\Sigma_{G}^{-1/2}-I_{r})]\]
and $\mathbb{W}(\pi)$ is an $r^{2}$ vector of independent Brownian
motions.\textit{\large{} }{\large \par}

(ii) Let $\Omega_{H}\equiv\mathrm{plim}_{T\to\infty}\var[T^{-1/2}\sum_{t=1}^{T}\vec(H_{0}^{\prime}g_{t}g_{t}^{\prime}H_{0}-I_{r})]$,
so using the definition of $R$ gives \begin{align*}
\Omega_{H} & =\mathrm{plim}_{T\to\infty}\var\left[\frac{1}{\sqrt{T}}\sum_{t=1}^{T}\vec[R^{\prime}(\eta_{t}\eta_{t}^{\prime}-I_{r})R]\right]\\
 & =\mathrm{plim}_{T\to\infty}\var\left[\frac{1}{\sqrt{T}}\sum_{t=1}^{T}(R^{\prime}\otimes R^{\prime})\vec(\eta_{t}\eta_{t}^{\prime}-I_{r})\right]\\
 & =(R^{\prime}\otimes R^{\prime})\Omega(R\otimes R).\end{align*}
Thus, the asymptotic distribution (\ref{eq:LR distribution trace})
can be expressed as \begin{align*}
 & \frac{1}{2\pi(1-\pi)}[\mathbb{W}(\pi)-\pi\mathbb{W}(1)]'\Omega[\mathbb{W}(\pi)-\pi\mathbb{W}(1)]\\
\overset{\mathrm{d}}{=} & \frac{1}{2\pi(1-\pi)}[\mathbb{W}(\pi)-\pi\mathbb{W}(1)]'(R^{\prime}\otimes R^{\prime})\Omega(R\otimes R)[\mathbb{W}(\pi)-\pi\mathbb{W}(1)]\\
= & \frac{1}{2\pi(1-\pi)}[\mathbb{W}(\pi)-\pi\mathbb{W}(1)]'\Omega_{H}[\mathbb{W}(\pi)-\pi\mathbb{W}(1)]\end{align*}
where $\overset{\mathrm{d}}{=}$ in the second line means equality
in distribution only and follows from the facts that $R\otimes R$
is orthonormal and $\mathbb{W}$ is a vector of independent Brownian
motions. Thus, we can use an estimator for $\Omega_{H}$ instead
of $\Omega$ to simulate the distribution. The HAC estimator in (\ref{eq:hac})
is consistent for $\Omega_{H}$ according to Theorem 2 of HI (2015). $\Box$
\textit{\large \vspace{1em}
}{\large \par}

{\bf Proof of Theorem \ref{thm2}:}
See the online appendix.
$\Box$
\textit{\large \vspace{1em}
}{\large \par}

{\bf Proof of Theorem \ref{thm power}:}

(i) Without loss of generality, we consider the case in which $B$ is singular
with $\mathrm{rank}(B)=r_{1}$ and $C$ is nonsingular under the alternative.
Given Assumption \ref{eig bound}, it follows that
\begin{align}
\mathrm{Prob}\left(\max_{[\epsilon T]\le k\le T_{1}}\rho_{j}(\hat{\Sigma}_{1}(k))\le\frac{c}{T}\right) & \to1,\label{eq:dbh prop1}\end{align}
for $j=r_{1}+1,...,r$ by Proposition 1 of DBH (2022) for some $c>0$. In addition,
for $k=T_{1}$ and $j=1,...,r_{1}$ we have, as $N,T\rightarrow \infty$, \[
\rho_{j}(\hat{\Sigma}_{1}(T_{1}))\to_{p}\rho_{j}(H_{0}^{\prime}B\Sigma_{F}B^{\prime}H_{0})\equiv v{}_{j}<\infty\]
by (\ref{eq:delta_NT_uniform}) and Assumption \ref{factors}. Thus,
for $k=T_{1}$ \begin{align}
-T_{1}\log(|\hat{\Sigma}_{1}(T_{1})|) & =-T_{1}\sum_{j=1}^{r_{1}}\log[\rho_{j}(\hat{\Sigma}_{1}(T_{1}))]-T_{1}\sum_{j=r_{1}+1}^{r}\log[\rho_{j}(\hat{\Sigma}_{1}(T_{1}))]\nonumber \\
 & \ge-T_{1}\sum_{j=1}^{r_{1}}\log(v_{j})-T_{1}(r-r_{1})\log(cT^{-1})\ w.p.a.1\nonumber \\
\frac{-T_{1}\log(|\hat{\Sigma}_{1}(T_{1})|)}{T\log T} & \ge\underbrace{\pi_{1}(r-r_{1})}_{\mathrm{leading\ term}}-\frac{\pi_{1}(r-r_{1})\log(c)}{\log T}-\frac{\pi_{1}\sum_{j=1}^{r_{1}}\log(v_{j})}{\log T}\ w.p.a.1\label{eq:klog(Sig1)}\end{align}
where $T_{1}/T\to\pi_{1}$ and the second line follows from (\ref{eq:dbh prop1}).
In addition, we have \begin{align}
\hat{\Sigma}_{2}(T_{1}) & =H_{0}^{\prime}C\Sigma_{F}C^{\prime}H_{0}+o_{p}(1)\label{eq:hat_Sig2}\end{align}
The RHS of (\ref{eq:hat_Sig2}) is positive definite because
$C$ is nonsingular, so \[
\frac{-(T-T_{1})\log(|\hat{\Sigma}_{2}(T_{1})|)}{T\log T}=-\frac{(1-\pi_{1})}{\log T}\sum_{j=1}^{r}\log\rho_{j}(H_{0}^{\prime}C\Sigma_{F}C^{\prime}H_{0})+o_{p}(1)\to_{p}0\]
is also dominated by the leading term in (\ref{eq:klog(Sig1)}). This
completes the proof for part (i).

(ii) For $k=T_{1}$, we have \[
\hat{\Sigma}_{2}(T_{1})=[I_{r}-\pi_{1}\hat{\Sigma}_{1}(T_{1})]/(1-\pi_{1}).\]
Consider the function \begin{align*}
 & \pi\log|\hat{\Sigma}_{1}(T_{1})|+(1-\pi)\log|(1-\pi)^{-1}[I_{r}-\pi\hat{\Sigma}_{1}(T_{1})]|\\
= & \pi\sum_{i=1}^{r}\log(\rho_{i})+(1-\pi)\sum_{i=1}^{r}\log\left(\frac{1-\pi\rho_{i}}{1-\pi}\right)\equiv\psi(\rho_{1},...,\rho_{r}),\end{align*}
where $\rho_{i}=\rho_{i}(\hat{\Sigma}_{1}(T_{1}))$ and $0<\rho_{i}<1/\pi$.
Note that \[
\partial\psi(\rho_{1},...,\rho_{r})/\partial\rho_{i}=\frac{\pi}{\rho_{i}}-(1-\pi)\frac{\pi}{1-\rho_{i}\pi}=\frac{\pi(1-\rho_{i})}{\rho_{i}(1-\rho_{i}\pi)},\]
so $\partial\psi(\rho_{i})/\partial\rho_{i}>0$ for $0<\rho_{i}<1$
and $\partial\psi(\rho_{i})/\partial\rho_{i}<0$ for $1<\rho_{i}<1/\pi$.
Thus, $\psi(\rho_{1},...,\rho_{r})$ takes its maximum equal to zero
when $\rho_{i}=1$ for $i=1,...,r$. Because $B\Sigma_{F}B^{\prime}\ne C\Sigma_{F}C^{\prime}$,
there must exist at least a $\rho_{i}\ne1$ for some $i$ so that
$\psi(\rho_{1},...,\rho_{r})<0$. Thus, \[
\sup\limits _{[\epsilon T]\le k\le[(1-\epsilon)T]}LR(k,\hat{G})\ge LR(T_{1},\hat{G})=-T\psi(\rho_{1},...,\rho_{r})>0,\]
which gives the desired result. $\Box$\textit{\large \vspace{1em}
}{\large \par}

{\bf Proof of Theorem \ref{thm mean break}:}

(i) Note that
\begin{align}
\tilde{\Sigma}_{1}(k) & =\frac{1}{k}\sum\limits _{t=1}^{k}\hat{g}_{t}\hat{g}_{t}^{\prime}-\bar{\hat{g}}_{1}\bar{\hat{g}}_{1}^{\prime}=I_{r}+\frac{1}{k}\sum\limits _{t=1}^{k}(\hat{g}_{t}\hat{g}_{t}^{\prime}-I_{r})-\bar{\hat{g}}_{1}\bar{\hat{g}}_{1}^{\prime}\nonumber \\
 & =I_{r}+O_{p}(T^{-1/2})+O_{p}(T^{-1}),\label{eq:tilde Sig1}\end{align}
where the $O_{p}(T^{-1/2})$ term follows from (\ref{eq:Sig1-Ir})
under the null hypothesis, the $O_{p}(T^{-1})$ term follows from
Assumption \ref{joint process} and (\ref{eq:need to prove}), and
both terms are uniform over $[\epsilon T]\le k\le[(1-\epsilon)T]$.

Thus, the second order Taylor expansion gives \begin{align*}
\log(|\tilde{\Sigma}_{1}(k)|) & =\tr\left[\frac{1}{k}\sum\limits _{t=1}^{k}(\hat{g}_{t}\hat{g}_{t}^{\prime}-I_{r})-\bar{\hat{g}}_{1}\bar{\hat{g}}_{1}^{\prime}\right]-\frac{1}{2}\tr\left(\left[\frac{1}{k}\sum\limits _{t=1}^{k}(\hat{g}_{t}\hat{g}_{t}^{\prime}-I_{r})-\bar{\hat{g}}_{1}\bar{\hat{g}}_{1}^{\prime}\right]^{2}\right)+o_{p}(T^{-1}),\end{align*}
where the $o_{p}(T^{-1})$ term follows from (\ref{eq:tilde Sig1})
and is uniform over $[\epsilon T]\le k\le[(1-\epsilon)T]$. Similarly,
\[
\log(|\tilde{\Sigma}_{2}(k)|)=\tr\left[\frac{1}{T-k}\sum\limits _{t=k+1}^{T}(\hat{g}_{t}\hat{g}_{t}^{\prime}-I_{r})-\bar{\hat{g}}_{2}\bar{\hat{g}}_{2}^{\prime}\right]-\frac{1}{2}\tr\left(\left[\frac{1}{T-k}\sum\limits _{t=k+1}^{T}(\hat{g}_{t}\hat{g}_{t}^{\prime}-I_{r})-\bar{\hat{g}}_{2}\bar{\hat{g}}_{2}^{\prime}\right]^{2}\right)+o_{p}(T^{-1}).\]
Thus, we have \begin{align}
k\log(|\tilde{\Sigma}_{1}(k)|) & =\tr\left(\sum\limits _{t=1}^{k}(\hat{g}_{t}\hat{g}_{t}^{\prime}-I_{r})\right)-k\bar{\hat{g}}_{1}^{\prime}\bar{\hat{g}}_{1}-\frac{k}{2}\tr\left(\left[\frac{1}{k}\sum\limits _{t=1}^{k}(\hat{g}_{t}\hat{g}_{t}^{\prime}-I_{r})-\bar{\hat{g}}_{1}\bar{\hat{g}}_{1}^{\prime}\right]^{2}\right)+o_{p}(1)\nonumber \\
 & =\tr\left(\sum\limits _{t=1}^{k}(\hat{g}_{t}\hat{g}_{t}^{\prime}-I_{r})\right)-k\bar{\hat{g}}_{1}^{\prime}\bar{\hat{g}}_{1}-\frac{k}{2}\tr\left(\left[\frac{1}{k}\sum\limits _{t=1}^{k}(\hat{g}_{t}\hat{g}_{t}^{\prime}-I_{r})\right]^{2}\right)\nonumber \\
 & +\frac{k}{2}\tr\left(\frac{2}{k}\sum\limits _{t=1}^{k}(\hat{g}_{t}\hat{g}_{t}^{\prime}-I_{r})\bar{\hat{g}}_{1}\bar{\hat{g}}_{1}^{\prime}\right)-\frac{k}{2}\tr(\bar{\hat{g}}_{1}\bar{\hat{g}}_{1}^{\prime})^{2}+o_{p}(1)\nonumber \\
 & =\tr\left(\sum\limits _{t=1}^{k}(\hat{g}_{t}\hat{g}_{t}^{\prime}-I_{r})\right)-k\bar{\hat{g}}_{1}^{\prime}\bar{\hat{g}}_{1}-\frac{k}{2}\tr\left(\left[\frac{1}{k}\sum\limits _{t=1}^{k}(\hat{g}_{t}\hat{g}_{t}^{\prime}-I_{r})\right]^{2}\right)+o_{p}(1),\label{eq:expansion of tilde_sig1}\end{align}
where the last equality follows from (\ref{eq:Sig1-Ir}) and Assumption
\ref{joint process}, and the $o_{p}(1)$ term is uniform over $[\epsilon T]\le k\le[(1-\epsilon)T]$.
Similarly, \[
(T-k)\log(|\tilde{\Sigma}_{2}(k)|)=\tr\left(\sum\limits _{t=k+1}^{T}(\hat{g}_{t}\hat{g}_{t}'-I_{r})\right)-(T-k)\bar{\hat{g}}_{2}^{\prime}\bar{\hat{g}}_{2}-\left(\frac{T-k}{2}\right)\tr\left(\left[\frac{1}{T-k}\sum\limits _{t=k+1}^{T}(\hat{g}_{t}\hat{g}_{t}'-I_{r})\right]^{2}\right)+o_{p}(1).\]
For the term $\bar{\hat{g}}_{1}$, we have \begin{align}
\frac{1}{\sqrt{T}}\sum\limits _{t=1}^{k}\hat{g}_{t} & =\frac{1}{\sqrt{T}}\sum\limits _{t=1}^{k}\hat{g}_{t}-\frac{k}{T}\left(\frac{1}{\sqrt{T}}\sum_{t=1}^{T}\hat{g}_{t}\right)\nonumber \\
 & =\frac{1}{\sqrt{T}}\sum\limits _{t=1}^{k}H^{\prime}g_{t}-\frac{k}{T}\left(\frac{1}{\sqrt{T}}\sum_{t=1}^{T}H^{\prime}g_{t}\right)+\frac{1}{\sqrt{T}}\sum\limits _{t=1}^{k}(\hat{g}_{t}-H^{\prime}g_{t})-\frac{k}{T}\left(\frac{1}{\sqrt{T}}\sum_{t=1}^{T}(\hat{g}_{t}-H^{\prime}g_{t})\right)\nonumber \\
 & =\frac{1}{\sqrt{T}}\sum\limits _{t=1}^{k}H^{\prime}g_{t}-\frac{k}{T}\left(\frac{1}{\sqrt{T}}\sum_{t=1}^{T}H^{\prime}g_{t}\right)+O_{p}(\sqrt{T}\delta_{NT}^{-2})\nonumber \\
 & =R^{\prime}\left[\frac{1}{\sqrt{T}}\sum\limits _{t=1}^{k}\eta_{t}-\frac{k}{T}\left(\frac{1}{\sqrt{T}}\sum_{t=1}^{T}\eta_{t}\right)\right]+O_{p}(\sqrt{T}\delta_{NT}^{-2})+o_{p}(1)\label{eq:bar_g1}\end{align}
with $\pi=k/T$, where the first equality uses the fact that $T^{-1}\sum\hat{g}_{t}=0$
because PCs are estimated from demeaned data, and the $O_{p}(\sqrt{T}\delta_{NT}^{-2})$
term follows from (\ref{eq:need to prove}) and is uniform over $k$. This also
implies that \begin{equation}
\frac{1}{\sqrt{T}}\sum\limits _{t=k+1}^{T}\hat{g}_{t}=-\frac{1}{\sqrt{T}}\sum\limits _{t=1}^{k}\hat{g}_{t}.\label{eq:bar_g2}\end{equation}
By (\ref{eq:bar_g2}) we have \begin{align}
k\bar{\hat{g}}_{1}'\bar{\hat{g}}_{1}+(T-k)\bar{\hat{g}}_{2}'\bar{\hat{g}}_{2} & \equiv\frac{T^{2}}{k(T-k)}\left(T^{-1/2}\sum_{t=1}^{k}\hat{g}_{t}\right)^{\prime}\left(T^{-1/2}\sum_{t=1}^{k}\hat{g}_{t}\right).\label{eq:mean term}\end{align}
In addition, \begin{align}
 & \frac{k}{2}\tr\left(\left[\frac{1}{k}\sum\limits _{t=1}^{k}(\hat{g}_{t}\hat{g}_{t}^{\prime}-I_{r})\right]^{2}\right)+\left(\frac{T-k}{2}\right)\tr\left(\left[\frac{1}{T-k}\sum\limits _{t=k+1}^{T}(\hat{g}_{t}\hat{g}_{t}^{\prime}-I_{r})\right]^{2}\right)\nonumber \\
= & \frac{T}{2k}\tr\left(\left[T^{-1/2}\sum\limits _{t=1}^{k}(\hat{g}_{t}\hat{g}_{t}^{\prime}-I_{r})\right]^{2}\right)+\frac{T}{2(T-k)}\tr\left(\left[T^{-1/2}\sum\limits _{t=k+1}^{T}(\hat{g}_{t}\hat{g}_{t}^{\prime}-I_{r})\right]^{2}\right)\nonumber \\
= & \frac{T^{2}}{2k(T-k)}\tr\left(\left[T^{-1/2}\sum\limits _{t=k+1}^{T}(\hat{g}_{t}\hat{g}_{t}^{\prime}-I_{r})\right]^{2}\right)\label{eq:variance term}\end{align}
 where the second equality follows from the first equation in (\ref{eq:gtgt2}).
By (\ref{eq:mean term}) and (\ref{eq:variance term}), we can obtain
\begin{eqnarray}
&&LR_{m}(k,\hat{G})\nonumber \\
 & =&\frac{T^{2}}{k(T-k)}\left[\begin{array}{c}
\frac{1}{\sqrt{T}}\sum_{t=1}^{k}\hat{g}_{t}\\
\frac{1}{\sqrt{2T}}\vec[\sum_{t=1}^{k}(\hat{g}_{t}\hat{g}_{t}'-I_{r})]\end{array}\right]^{\prime}\left[\begin{array}{c}
\frac{1}{\sqrt{T}}\sum_{t=1}^{k}\hat{g}_{t}\\
\frac{1}{\sqrt{2T}}\vec[\sum_{t=1}^{k}(\hat{g}_{t}\hat{g}_{t}'-I_{r})]\end{array}\right]+o_{p}(1)\nonumber \\
 & =&\frac{T^{2}}{k(T-k)}\left[\begin{array}{c}
\left[\frac{1}{\sqrt{T}}\sum\limits _{t=1}^{k}\eta_{t}-\frac{k}{T}\left(\frac{1}{\sqrt{T}}\sum_{t=1}^{T}\eta_{t}\right)\right]\\
\frac{1}{\sqrt{2T}}\vec\left[\sum\limits _{t=1}^{k}(\eta_{t}\eta_{t}^{\prime}-I_{r})-\frac{k}{T}\sum\limits _{t=1}^{T}(\eta_{t}\eta_{t}^{\prime}-I_{r})\right]\end{array}\right]^{\prime}\left[\begin{array}{c}
\left[\frac{1}{\sqrt{T}}\sum\limits _{t=1}^{k}\eta_{t}-\frac{k}{T}\left(\frac{1}{\sqrt{T}}\sum_{t=1}^{T}\eta_{t}\right)\right]\\
\frac{1}{\sqrt{2T}}\vec\left[\sum\limits _{t=1}^{k}(\eta_{t}\eta_{t}^{\prime}-I_{r})-\frac{k}{T}\sum\limits _{t=1}^{T}(\eta_{t}\eta_{t}^{\prime}-I_{r})\right]\end{array}\right]\nonumber \\
&&+o_{p}(1)\nonumber \\
 & \Rightarrow&\frac{[\mathbb{U}(\pi)-\pi\mathbb{U}(1)]'\bar{\Omega}[\mathbb{U}(\pi)-\pi\mathbb{U}(1)]}{\pi(1-\pi)},\label{eq:joint distribution for single mean change}\end{eqnarray}
where the second equality follows from $RR'=I_{r}$, (\ref{eq:gtgt-Ir expansion})
and (\ref{eq:bar_g1}), and the weak convergence follows from Assumption
\ref{joint process}.

(ii) Under the additional assumption that $\{g_{t}\}$ is uncorrelated
with $\{g_{t}g_{t}'-\Sigma_{G}\}$, (\ref{eq:mean term}) reduces
to \begin{align*}
k\bar{\hat{g}}_{1}^{\prime}\bar{\hat{g}}_{1}+(T-k)\bar{\hat{g}}_{2}^{\prime}\bar{\hat{g}}_{2} & =\frac{T^{2}}{k(T-k)}\left[\frac{1}{\sqrt{T}}\sum\limits _{t=1}^{k}\eta_{t}-\frac{k}{T}\left(\frac{1}{\sqrt{T}}\sum_{t=1}^{T}\eta_{t}\right)\right]^{\prime}\left[\frac{1}{\sqrt{T}}\sum\limits _{t=1}^{k}\eta_{t}-\frac{k}{T}\left(\frac{1}{\sqrt{T}}\sum_{t=1}^{T}\eta_{t}\right)\right]+o_{p}(1)\\
 & \Rightarrow\frac{1}{\pi(1-\pi)}[\mathcal{V}(\pi)-\pi\mathcal{V}(1)]^{\prime}[\mathcal{V}(\pi)-\pi\mathcal{V}(1)]\\
 & \overset{\mathrm{d}}{=}\frac{1}{\pi(1-\pi)}[\mathbb{W}_{1}(\pi)-\pi\mathbb{W}_{1}(1)]'\Omega_{1}[\mathbb{W}_{1}(\pi)-\pi\mathbb{W}_{1}(1)],\end{align*}
where $\mathbb{W}_{1}$ is an $r$-dimensional independent Brownian
motion process and $\Omega_{1}=\mathrm{plim}_{T\to\infty}\var\left(T^{-1/2}\sum_{t=1}^{k}\eta_{t}\right).$
Therefore, together with the result in (\ref{eq:LR distribution trace}),
the LR statistic has the following limiting distribution
\begin{eqnarray*}
&&LR_{m}(k,\hat{G}) \\ &=&k\bar{\hat{g}}_{1}^{\prime}\bar{\hat{g}}_{1}+(T-k)\bar{\hat{g}}_{2}^{\prime}\bar{\hat{g}}_{2}+\frac{k}{2}\tr\left(\left[\frac{1}{k}\sum\limits _{t=1}^{k}(\hat{g}_{t}\hat{g}_{t}'-I_{r})\right]^{2}\right)+\left(\frac{T-k}{2}\right)\tr\left(\left[\frac{1}{T-k}\sum\limits _{t=k+1}^{T}(\hat{g}_{t}\hat{g}_{t}'-I_{r})\right]^{2}\right)+o_{p}(1)\\
&\Rightarrow&\frac{[\mathbb{W}_{1}(\pi)-\pi\mathbb{W}_{1}(1)]'\Omega_{1}[\mathbb{W}_{1}(\pi)-\pi\mathbb{W}_{1}(1)]}{\pi(1-\pi)}+\frac{1}{2}\frac{[\mathbb{W}(\pi)-\pi\mathbb{W}(1)]'\Omega[\mathbb{W}(\pi)-\pi\mathbb{W}(1)]}{\pi(1-\pi)}.\end{eqnarray*}
Taking the supreme over $k$ yields the desired distribution. $\Box$\textit{\large \vspace{1em}
}{\large \par}

{\bf Proof of Theorem \ref{thm multi breaks}:}

Following similar steps in (\ref{eq:gtgt-Ir expansion}), we can obtain
\begin{eqnarray}
T^{-1/2}\sum\limits _{t=k_{j}+1}^{k_{j+1}}(\hat{g}_{t}\hat{g}_{t}'-I_{r}) & = & H^{\prime}\left(\frac{1}{\sqrt{T}}\sum\limits _{t=k_{j}+1}^{k_{j+1}}g_{t}g_{t}^{\prime}-\frac{k_{j+1}-k_{j}}{T}\frac{1}{\sqrt{T}}\sum\limits _{t=1}^{T}g_{t}g_{t}^{\prime}\right)H+\sqrt{T}O_{p}(\delta_{NT}^{-2})\nonumber \\
 & = & H^{\prime}\left(\frac{1}{\sqrt{T}}\sum\limits _{t=k_{j}+1}^{k_{j+1}}(g_{t}g_{t}^{\prime}-\Sigma_{G})-\frac{k_{j+1}-k_{j}}{T}\frac{1}{\sqrt{T}}\sum\limits _{t=1}^{T}(g_{t}g_{t}^{\prime}-\Sigma_{G})\right)H+\sqrt{T}O_{p}(\delta_{NT}^{-2})\nonumber \\
 & = & R^{\prime}\left(\frac{1}{\sqrt{T}}\sum\limits _{t=k_{j}+1}^{k_{j+1}}(\eta_{t}\eta_{t}^{\prime}-I_{r})-\frac{k_{j+1}-k_{j}}{T}\frac{1}{\sqrt{T}}\sum\limits _{t=1}^{T}(\eta_{t}\eta_{t}^{\prime}-I_{r})\right)R+\sqrt{T}O_{p}(\delta_{NT}^{-2})\nonumber \\
 &&+o_{p}(1),\label{eq:term in m breaks}\end{eqnarray}
where the $O_{p}(\delta_{NT}^{-2})$ and $o_{p}(1)$ terms are uniform
in $k$.

(\ref{eq:term in m breaks}) implies that $\hat{\Sigma}_{j+1}-I_{r}$ is
uniformly $O_{p}(T^{-1/2})$, so following arguments similar to those used
in (\ref{eq:Sig1-Ir}) and (\ref{eq:each term in LR}), we have \begin{align}
\log(|\hat{\Sigma}_{j+1}|) & =\tr[\hat{\Sigma}_{j+1}-I_{r}]-\frac{1}{2}\tr((\hat{\Sigma}_{j+1}-I_{r})(\hat{\Sigma}_{j+1}-I_{r})+o_{p}(T^{-1})\nonumber \\
(k_{j+1}-k_{j})\log(|\hat{\Sigma}_{j+1}|) & =\tr\left(\sum\limits _{t=k_{j}+1}^{k_{j+1}}(\hat{g}_{t}\hat{g}_{t}^{\prime}-I_{r})\right)-\frac{k_{j+1}-k_{j}}{2}\tr\left(\left[\frac{1}{k_{j+1}-k_{j}}\sum\limits _{t=k_{j}+1}^{k_{j+1}}(\hat{g}_{t}\hat{g}_{t}^{\prime}-I_{r})\right]^{2}\right)+o_{p}(1),\label{eq:taylor expansion m breaks}\end{align}
where the $o_{p}(T^{-1})$ and $o_{p}(1)$ terms are uniform over
$[\epsilon T]\le k\le[(1-\epsilon)T]$ for $\epsilon\in(0,1)$. Because
$T^{-1}\sum_{t=1}^{T}\hat{g}_{t}\hat{g}_{t}^{\prime}=I_{r}$, we have\begin{align}
LR(k_{1},...,k_{m},\hat{G}) & =\frac{1}{2}\sum_{j=0}^{m}(k_{j+1}-k_{j})\tr\left(\left[\frac{1}{k_{j+1}-k_{j}}\sum\limits _{t=k_{j}+1}^{k_{j+1}}(\hat{g}_{t}\hat{g}_{t}^{\prime}-I_{r})\right]^{2}\right)+o_{p}(1)\nonumber \\
 & =\frac{1}{2}\sum_{j=0}^{m}\frac{T}{k_{j+1}-k_{j}}\tr\left(\left[T^{-1/2}\sum\limits _{t=k_{j}+1}^{k_{j+1}}(\hat{g}_{t}\hat{g}_{t}^{\prime}-I_{r})\right]^{2}\right)+o_{p}(1)\nonumber \\
 & =\frac{1}{2}\sum_{j=0}^{m}\frac{T}{k_{j+1}-k_{j}}\tr\left[\left(\frac{1}{\sqrt{T}}\sum\limits _{t=k_{j}+1}^{k_{j+1}}(\eta_{t}\eta_{t}^{\prime}-I_{r})-\frac{k_{j+1}-k_{j}}{T}\frac{1}{\sqrt{T}}\sum\limits _{t=1}^{T}(\eta_{t}\eta_{t}^{\prime}-I_{r})\right)^{2}\right]+o_{p}(1)\nonumber \\
 & \Rightarrow \sum_{j=0}^{m}\frac{1}{2(\pi_{j+1}-\pi_{j})}\tr\left([\zeta(\pi_{j+1})-\zeta(\pi_{j})-(\pi_{j+1}-\pi_{j})\zeta(1)]^{2}\right)
 \label{eq:m breaks without mean change distribution}
 \end{align}
where the third line follows from (\ref{eq:term in m breaks}) and the
fact that $RR^{\prime}=I_{r}$, and the weak convergence follows from
Assumption \ref{limit_distribution2}. The result in (\ref{eq:m breaks without mean change distribution}) can be written as
\begin{align*}
 & \vec\left(\zeta(\pi_{j+1})-\zeta(\pi_{j})-(\pi_{j+1}-\pi_{j})\zeta(1)\right)'\vec\left(\zeta(\pi_{j+1})-\zeta(\pi_{j})-(\pi_{j+1}-\pi_{j})\zeta(1)\right)\nonumber \\
\overset{\mathrm{d}}{=} &[\mathbb{W}(\pi_{j+1})-\mathbb{W}(\pi_{j})-(\pi_{j+1}-\pi_{j})\mathbb{W}(1)]'\Omega[\mathbb{W}(\pi_{j+1})-\mathbb{W}(\pi_{j})-(\pi_{j+1}-\pi_{j})\mathbb{W}(1)],\label{eq:LR distribution trace}\end{align*}
where $\overset{\mathrm{d}}{=}$ means equality in distribution only,
\[
\Omega=\mathrm{plim}_{T\to\infty}\var[\frac{1}{\sqrt{T}}\sum_{t=1}^{T}\vec(\Sigma_{G}^{-1/2}g_{t}g_{t}^{\prime}\Sigma_{G}^{-1/2}-I_{r})]\]
and $\mathbb{W}(\pi)$ is an $r^{2}$ vector of independent Brownian
motions. Thus, we have
\begin{align*}
&LR(k_{1},...,k_{m},\hat{G}) \\
& \Rightarrow  \sum_{j=0}^{m}\frac{1}{2(\pi_{j+1}-\pi_{j})} [\mathbb{W}(\pi_{j+1})-\mathbb{W}(\pi_{j})-(\pi_{j+1}-\pi_{j})\mathbb{W}(1)]'\Omega[\mathbb{W}(\pi_{j+1})-\mathbb{W}(\pi_{j})-(\pi_{j+1}-\pi_{j})\mathbb{W}(1)]
\\
&=\sum_{j=0}^{m}\frac{1}{2(\pi_{j+1}-\pi_{j})}  \|  \Omega^{1/2}[\tilde{\mathbb{B}}(\pi_{j+1})-\tilde{\mathbb{B}}(\pi_{j})]   \|^2
 \end{align*}
where $\tilde{\mathbb{B}}(\pi)=\mathbb{W}(\pi)-\pi \mathbb{W}(1)$ is a Brownian bridge process.

$\Box$\textit{\large \vspace{1em}
}{\large \par}

{\bf Proof of Theorem \ref{allow_mean_change_distribution_theorem_f_ff_sigma_correlated}:}

Following the steps in (\ref{eq:bar_g1}), we can obtain

\begin{equation}
\frac{1}{\sqrt{T}}\sum\limits _{t=k_{j}+1}^{k_{j+1}}\hat{g}_{t}=R^{\prime}\left[\frac{1}{\sqrt{T}}\sum\limits _{t=k_{j}+1}^{k_{j+1}}\eta_{t}-\frac{k_{j+1}-k_{j}}{T}\left(\frac{1}{\sqrt{T}}\sum_{t=1}^{T}\eta_{t}\right)\right]+O_{p}(\sqrt{T}\delta_{NT}^{-2})+o_{p}(1).\label{eq:bar_g1_multiple}\end{equation}
Similar to (\ref{eq:expansion of tilde_sig1}), we can obtain\begin{align*}
(k_{j+1}-k_{j})\log(|\tilde{\Sigma}_{j+1}|) & =\tr\left(\sum\limits _{t=k_{j}+1}^{k_{j+1}}(\hat{g}_{t}\hat{g}_{t}^{\prime}-I_{r})\right)-(k_{j+1}-k_{j})\bar{\hat{g}}_{j+1}^{\prime}\bar{\hat{g}}_{j+1}\\
 & -\frac{k_{j+1}-k_{j}}{2}\tr\left(\left[\frac{1}{k_{j+1}-k_{j}}\sum\limits _{t=k_{j}+1}^{k_{j+1}}(\hat{g}_{t}\hat{g}_{t}^{\prime}-I_{r})\right]^{2}\right)+o_{p}(1),\end{align*}
which has an extra term $(k_{j+1}-k_{j})\bar{\hat{g}}_{j+1}^{\prime}\bar{\hat{g}}_{j+1}$
compared to (\ref{eq:taylor expansion m breaks}). Thus, \begin{eqnarray*}
LR_{m}(k_{1},...,k_{m},\hat{G}) & = & \sum_{j=0}^{m}(k_{j+1}-k_{j})\bar{\hat{g}}_{j+1}^{\prime}\bar{\hat{g}}_{j+1}+\frac{(k_{j+1}-k_{j})}{2}\tr\left(\left[\frac{1}{k_{j+1}-k_{j}}\sum\limits _{t=k_{j}+1}^{k_{j+1}}(\hat{g}_{t}\hat{g}_{t}^{\prime}-I_{r})\right]^{2}\right)+o_{p}(1)\\
 & = & \sum_{j=0}^{m}\frac{T}{k_{j+1}-k_{j}}\left[\begin{array}{c}
\frac{1}{\sqrt{T}}\sum\limits _{t=k_{j}+1}^{k_{j+1}}\hat{g}_{t}\\
\frac{1}{\sqrt{2T}}\vec[\sum\limits _{t=k_{j}+1}^{k_{j+1}}(\hat{g}_{t}\hat{g}_{t}'-I_{r})]\end{array}\right]^{\prime}\left[\begin{array}{c}
\frac{1}{\sqrt{T}}\sum\limits _{t=k_{j}+1}^{k_{j+1}}\hat{g}_{t}\\
\frac{1}{\sqrt{2T}}\vec[\sum\limits _{t=k_{j}+1}^{k_{j+1}}(\hat{g}_{t}\hat{g}_{t}'-I_{r})]\end{array}\right]+o_{p}(1)\\
 & = & \sum_{j=0}^{m}\frac{T}{k_{j+1}-k_{j}}\left\Vert \begin{array}{c}
\frac{1}{\sqrt{T}}\sum\limits _{t=k_{j}+1}^{k_{j+1}}\eta_{t}-\frac{k_{j+1}-k_{j}}{T}\left(\frac{1}{\sqrt{T}}\sum_{t=1}^{T}\eta_{t}\right)\\
\frac{1}{\sqrt{2T}}\vec[\sum\limits _{t=k_{j}+1}^{k_{j+1}}(\eta_{t}\eta_{t}^{\prime}-I_{r})-\frac{k_{j+1}-k_{j}}{T}\sum\limits _{t=1}^{T}(\eta_{t}\eta_{t}^{\prime}-I_{r})]\end{array}\right\Vert ^{2}+o_{p}(1),\end{eqnarray*}
where the last equality follows from (\ref{eq:bar_g1_multiple}) and
(\ref{eq:term in m breaks}). Under Assumption \ref{joint process} and
arguments similar to those used in (\ref{eq:joint distribution for single mean change})
and (\ref{eq:m breaks without mean change distribution}), we have
\begin{align*}
LR_{m}(k_{1},...,k_{m},\hat{G}) & \Rightarrow\sum\limits _{j=0}^{m}\frac{1}{\pi_{j+1}-\pi_{j}}\left\Vert \bar{\Omega}^{1/2}[U(\pi_{j+1})-U(\pi_{j})-(\pi_{j+1}-\pi_{j})U(1)]\right\Vert ^{2}\\
 & =\sum\limits _{j=0}^{m}\frac{\|\bar{\Omega}^{1/2}[\mathbb{B}(\pi_{j+1})-\mathbb{B}(\pi_{j})]\|^{2}}{\pi_{j+1}-\pi_{j}}\end{align*}
where $\mathbb{B}(\pi)=U(\pi)-\pi U(1)$ is a Brownian bridge process. $\Box$\textit{\large \vspace{1em}
}{\large \par}

\end{document}